\pdfoutput=1 
\documentclass{article}

\usepackage{algorithm}
\usepackage[margin=30mm,includehead,includefoot]{geometry}
\usepackage[noend]{algpseudocode}
\usepackage{amsmath,amssymb,amsthm,amscd,color,comment}
\usepackage{autobreak}
\usepackage[dvipsnames]{xcolor}
\usepackage{url}
\usepackage{jheppub}
\usepackage{empheq}
\usepackage{graphicx}
\usepackage{subfig}
\usepackage{parskip} 
\usepackage{mathtools}
\usepackage{enumitem}
\usepackage{nicematrix,tikz}

\usepackage{xpatch}
\makeatletter
\AtBeginDocument{\xpatchcmd{\@thm}{\thm@headpunct{.}}{\thm@headpunct{}}{}{}}
\makeatother

\newcommand{\brk}[1]{(#1)}
\newcommand{\lrbrk}[1]{\left(#1\right)}
\newcommand{\bigbrk}[1]{\bigl(#1\bigr)}
\newcommand{\Bigbrk}[1]{\Bigl(#1\Bigr)}

\newcommand{\sbrk}[1]{[#1]}

\newcommand{\Bigsbrk}[1]{\Bigl[#1\Bigr]}

\newcommand{\brc}[1]{\{#1\}}

\newcommand{\bigbrc}[1]{\bigl\{#1\bigr\}}

\newcommand{\abs}[1]{|#1|}
\newcommand{\vev}[1]{\langle #1\rangle}
\newcommand{\bigvev}[1]{\big\langle #1\big\rangle}
\newcommand{\supbrk}[1]{^{\brk{#1}}}
\newcommand{\brkbf}[1]{{\brk{\mathbf{#1}}}}
\newcommand{\supbbf}[1]{^{\brk{\mathbf{#1}}}}

\newcommand{\ket}[1]{| #1 \rangle}
\DeclarePairedDelimiterX\braket[2]{\langle}{\rangle}{#1 \>\delimsize\vert\> #2}

\newcommand{\dd}{\mathrm{d}}
\newcommand{\dlog}{\dd \log}
\newcommand{\defas}{:=}
\newcommand{\safed}{=:}

\newcommand{\Complex}{\mathbb{C}}
\newcommand{\CP}{\mathbb{C}\mathrm{P}}

\newcommand{\Poles}{\mathcal{P}}

\newcommand{\projPhi}{\varphi}
\newcommand{\Baikov}{\mathcal{B}}
\newcommand{\Res}{{\rm Res}}
\newcommand{\Cmat}{{\bf C}}
\newcommand{\GCD}{\mathrm{GCD}}
\newcommand{\LCM}{\mathrm{LCM}}
\newcommand{\LT}{\mathrm{LT}}

\newcommand{\cS}{\mathcal{S}}
\newcommand{\cX}{\mathcal{X}}
\newcommand{\rmH}{\mathrm{H}}
\newcommand{\Dim}{\mathrm{dim}}
\newcommand{\layer}[1]{^{\brk{#1}}}
\newcommand{\MIs}{e}

\makeatletter
    \newcommand*\bigcdot{\mathpalette\bigcdot@{1}}
    \newcommand*\smtimes{\mathpalette\smtimes@{.7}}
    \newcommand*\bigcdot@[2]{\mathbin{\vcenter{\hbox{\scalebox{#2}{$\m@th#1\bullet$}}}}}
    \newcommand*\smtimes@[2]{\mathbin{\vcenter{\hbox{\scalebox{#2}{$\m@th#1\times$}}}}}
\makeatother

\newcommand{\soft}[1]{\textsc{#1}}

\newcommand{\filename}[1]{\texttt{#1}}
\newcommand{\code}[1]{\texttt{#1}}

\newcommand{\namedref}[2]{\hyperref[#2]{#1~\ref*{#2}}}
\newcommand{\secref}[1]{\namedref{Section}{#1}}

\newcommand{\appref}[1]{\namedref{Appendix}{#1}}

\newcommand{\tabref}[1]{\namedref{Table}{#1}}

\newcommand{\figref}[1]{\namedref{Figure}{#1}}

\makeatletter
\def\mr@ignsp#1 {\ifx\:#1\@empty\else #1\expandafter\mr@ignsp\fi}%
\newcommand{\multiref}[1]{\begingroup
\xdef\mr@no@sparg{\expandafter\mr@ignsp#1 \: }%
\def\mr@comma{}%
\@for\mr@refs:=\mr@no@sparg\do{\mr@comma\def\mr@comma{,\,}\ref{\mr@refs}}%
\endgroup}
\makeatother
\renewcommand{\eqref}[1]{(\multiref{#1})}

\newcommand{\be}{\begin{equation}}
\newcommand{\ee}{\end{equation}}
\newcommand{\bea}{\begin{eqnarray}}
\newcommand{\eea}{\end{eqnarray}}
\newcommand{\bei}{\begin{itemize}}
\newcommand{\eei}{\end{itemize}}

\makeatletter
\newlength{\apb@width}
\newcommand{\autoparbox}[2][c]{\settowidth{\apb@width}{#2}\parbox[#1]{\apb@width}{#2}}
\newcommand{\includegraphicsbox}[2][]{\autoparbox{\includegraphics[#1]{#2}}}
\makeatother

\definecolor{green1}{HTML}{244819}
\definecolor{cyan1}{HTML}{37cdaa}
\definecolor{blue1}{HTML}{5d7ac4}
\definecolor{red1}{HTML}{d0482a}
\definecolor{purple1}{HTML}{845ea8}
\definecolor{orange1}{HTML}{e07229}

\usepackage{etoolbox}
\makeatletter
    \newcommand\newsubsupcommand[4]{\newcommand#1{#2\sc@subp{#3}{#4}}}
    \def\sc@subp#1#2{%
        \let\sc@subflag\undefinded%
        \let\sc@supflag\undefinded%
        \def\sc@thesub{#1}%
        \def\sc@thesup{#2}%
        \sc@proc%
    }%
    \def\sc@proc{%
        \@ifnextchar{_}{\def\sc@subflag{}\sc@mergesubs}{%
            \@ifnextchar{^}{\def\sc@supflag{}\sc@mergesups}{
                \ifdef{\sc@subflag}{}{_{\sc@thesub}}%
                \ifdef{\sc@supflag}{}{^{\sc@thesup}}%
            }%
        }%
    }%
    \def\sc@mergesubs#1#2{_{\sc@thesub#2}\sc@proc}%
    \def\sc@mergesups#1#2{^{\sc@thesup#2}\sc@proc}%
\makeatother

\newsubsupcommand{\phiL}{\varphi}{L}{}
\newsubsupcommand{\phiR}{\varphi}{R}{}
\newsubsupcommand{\psiL}{\psi}{L}{}
\newsubsupcommand{\psiR}{\psi}{R}{}

\title{Intersection Numbers, Polynomial Division and Relative Cohomology}

\author[a,b,c]{Giacomo Brunello,}
\author[a,b,d]{Vsevolod Chestnov,}
\author[a,b]{Giulio Crisanti,} 
\author[e]{Hjalte Frellesvig,}
\author[a,b]{Manoj K. Mandal,}
\author[a,b]{Pierpaolo Mastrolia}

\newcommand{\unipd}{Dipartimento di Fisica e Astronomia, Universit\`a degli Studi di Padova,
Via Marzolo 8, I-35131 Padova, Italy.}

\newcommand{\pdinfn}{INFN, Sezione di Padova,
Via Marzolo 8, I-35131 Padova, Italy.}

\newcommand{\bologna}{
    Dipartimento di Fisica e Astronomia, Universit\`a di Bologna
    e INFN, Sezione di Bologna,
    via Irnerio 46, I-40126 Bologna, Italy.
}

\affiliation[a]{\unipd}
\affiliation[b]{\pdinfn}
\affiliation[c]{Institut de Physique Théorique, CEA, CNRS, Université Paris-Saclay, F–91191 Gif-sur-Yvette cedex, France}
\affiliation[d]{\bologna}
\affiliation[e]{Niels Bohr International Academy, University of Copenhagen,
Blegdamsvej 17, 2100 K{\o}benhavn {\O}, Denmark}

\emailAdd{giacomo.brunello@phd.unipd.it}
\emailAdd{vsevolod.chestnov@unibo.it}
\emailAdd{giulioeugenio.crisanti@phd.unipd.it}
\emailAdd{hjalte.frellesvig@nbi.ku.dk}
\emailAdd{manojkumar.mandal@pd.infn.it}
\emailAdd{pierpaolo.mastrolia@unipd.it}

\abstract{
We present a simplification of the recursive algorithm for the evaluation of intersection numbers for differential $n$-forms, by combining the advantages emerging from the choice of delta-forms as generators of relative twisted cohomology groups and the polynomial division technique, recently proposed in the literature.
We show that delta-forms capture the leading behaviour of the intersection numbers in presence of evanescent analytic regulators, whose use is, therefore, bypassed.
This simplified algorithm is applied to derive the complete decomposition of two-loop planar and non-planar Feynman integrals in terms of a master integral basis. More generally, it can be applied to derive relations among twisted period integrals, relevant for physics and mathematical studies.
}

\begin{document}

\maketitle

\section{Introduction}
\label{sec:intro}

The intersection number between differential $n$-forms
\cite{matsumoto1994,matsumoto1998,OST2003,doi:10.1142/S0129167X13500948,goto2015,goto2015b,Yoshiaki-GOTO2015203,Mizera:2017rqa,matsubaraheo2019algorithm, Ohara98intersectionnumbers, https://doi.org/10.48550/arxiv.2006.07848, https://doi.org/10.48550/arxiv.2008.03176, https://doi.org/10.48550/arxiv.2104.12584}
is an elementary quantity that rules the vector space properties of regulated
(twisted period) integrals. The respective integrands are defined through the product of
the \textit{twist}, a regulating multivalued function that vanishes at the
boundary of the integration domain, and a differential $n$-form.
Acting as an inner product, the intersection number yields the decomposition of the
differential forms into a basis of forms that generate the twisted de Rham
cohomology group, a vector space defined as the quotient space of the closed
forms {\it modulo} the exact forms
\cite{Mastrolia:2018uzb,Frellesvig:2019kgj,Frellesvig:2019uqt,Frellesvig:2020qot}.
Linear and quadratic
relations among the elements of the vector space, as well as differential and
difference equations for them, can be derived using intersection numbers.
The translation of these identities to their respective integral formulations
is then straightforward.

The algebraic properties of Feynman integrals, central
objects of study in perturbative classical and quantum field theory, motivated
us to develop the intersection theory framework. In that framework the linear
relations derived by intersection numbers are equivalent to the well-known
integration-by-parts identities \brk{IBPs}
\cite{Tkachov:1981wb,Chetyrkin:1981qh},
used to derive the integrals decomposition in terms of an independent set of master integrals (MIs), as well as for the differential and difference equations obeyed by them.
The same formalism can be applied to a wider class of functions, such as
Aomoto-Gelfand and Euler-Mellin integrals, as well as Gelfand-Kapranov-Zelevinsky
hypergeometric systems
\cite{GKZ-1989,GKZ-Euler-1990,Chestnov:2022alh} which embed Feynman integrals
as special restrictions \cite{Chestnov:2023kww} (see also \cite{Agostini:2022cgv,Matsubara-Heo:2023ylc}).
Beyond IBP reductions, twisted co-homology finds applications in many other relevant areas of Physics and Mathematics: the formalism has been applied in the
construction of the canonical bases for Feynman integrals~\cite{Chen:2020uyk, Chen:2022lzr, Chen:2023kgw,Giroux:2022wav} (in presence of generalised polylogarithms as well as of elliptic functions),
correlator functions in quantum field theory and lattice gauge theory
\cite{Weinzierl:2020gda,Weinzierl:2020nhw,Cacciatori:2022mbi,Gasparotto:2022mmp,Gasparotto:2023roh}
(in perturbative and non-perturbative approaches),
orthogonal polynomials, quantum mechanical
matrix elements, Witten-Kontsevich tau-functions \cite{Cacciatori:2022mbi},
cosmological correlators \cite{De:2023xue},
representation of Feynman integrals as single-valued hypergeometric functions~\cite{Duhr:2023bku},
to name a few\footnote{
    More examples can be found in the recent
    theses~\cite{MassiddaThesis,Mattiazzi:2022zbo,Gasparotto:2023cdl}, in the lecture notes~\cite{Matsubara-Heo:2023ylc}
    and the in the reviews~\cite{Cacciatori:2021nli,Weinzierl:2022eaz}.
    For related works, see also the recent \cite{Artico:2023jrc}.
}.

There exist several methods for the calculation of intersection numbers using the
twisted version of Stokes' theorem \cite{cho1995}. In the special case of
logarithmic $n$-forms they can be computed via the algorithms proposed in
\cite{matsumoto1998,Mizera:2017rqa,Mizera:2019vvs,Matsubara-Heo:2021dtm}.
In the more general situation of meromorphic $n$-forms,
the calculation of the intersection number can
proceed according to the so-called \textit{recursive  approach}, as proposed in
\cite{Mizera:2019gea,Frellesvig:2019uqt,Frellesvig:2020qot}, elaborating on
\cite{Ohara98intersectionnumbers}, which, exploiting the concept of {\it fibration},
maps the (evaluation of) intersection numbers for $n$-forms, into an ordered sequence of (evaluations of) intersection numbers for 1-forms, that are computed by considering one (integration) variable at a time.
This method can be simplified by exploiting the invariance of the representative of the cohomology classes and by avoiding the use of algebraic extensions \cite{Weinzierl:2020xyy},
and its application range can be extended \cite{Caron-Huot:2021xqj,Caron-Huot:2021iev} also to the case of the \textit{relative twisted cohomology} \cite{matsumoto2018relative},
that deals with singularities of the integrand not regulated by the twist.
An interesting improvement of the evaluation procedure, that
makes use of the idea of avoiding polynomial factorisation and algebraic extensions,
has been recently proposed in \cite{Fontana:2023amt}, by introducing the \textit{polynomial series expansion} technique, and
an efficient treatment of the analytic regulators,
combined with the advantages of modular arithmetic over the finite fields~\cite{Peraro:2016wsq,Peraro:2019svx}.

Also, we have recently proposed a new algorithm for computing the intersection number of twisted $n$-forms, based on a multivariate version of Stokes' theorem,
which requires the solution of a {\it higher-order partial differential
equation} and the evaluation of {\it multivariate residues} \cite{Chestnov:2022xsy}, which is a natural
generalization of the original algorithm \cite{cho1995, matsumoto1998} that avoids the fibration procedure.

Let us finally mention that an alternative method for evaluation of
intersection numbers, which is based on the solution of the \textit{secondary
equation} built from the \textit{Pfaffian system} of differential equations for
the generators of the cohomology group
\cite{Matsubara-Heo-Takayama-2020b,matsubaraheo2019algorithm}, in combination
with an efficient algorithm for construction of such systems by means of the
Macaulay matrix has recently been proposed in \cite{Chestnov:2022alh}.

The previous activities show that the development of the optimal algorithms for
evaluating intersection numbers for meromorphic twisted $n$-forms is an open
problem of common interest for mathematicians and physicists.

The classification of dimensionally regulated Feynman integrals as twisted period integrals becomes manifest when, instead of the canonical momentum-space representation,
parametric representations of the integrals are adopted \cite{Lee:2013hzt,Mastrolia:2018uzb}.
Additional analytic regularisation \cite{Speer:1969sjv, Speer1971} is often required to regulate the behaviour of inverse powers of the integration variables that arise upon the change of variables, from the loop momenta to the new set of $n$ integration variables.
Accordingly, auxiliary non-integer regulating exponents, hereby dubbed ${\it regulators}$, are introduced in the definition of the twist \cite{Frellesvig:2017aai},
which are later set to zero at the end of the calculation of the coefficients of the MIs decomposition.
Therefore, intersection numbers acquire a dependence on the evanescent
regulators, beside the expected dependence on external kinematic invariants,
masses, and the continuous space-time dimensions $d$.  For that reason the
regulators affect the load of the evaluation algorithm, which would be much
lighter if they were absent.

In the case of Euler-Mellin integrals and GKZ hypergeometric functions, non-integer exponents are considered
{\it ab-initio} in the integral representation, and, in these cases, the intersection numbers depend on them, as well as on the multiplicative factors of the monomials (product of integration variables) that appear in the twist - which can be considered as external variables.

In this work, we investigate the analytic behaviour of the intersection numbers
of 1-forms on one evanescent regulator, and show that the coefficients of the
MIs decomposition can be computed just using the leading term (LT) of the
Laurent series expansion of the intersection numbers, as it was remarkably
observed, by other means, in \cite{Fontana:2023amt}.  Moreover, we show that
the expression of the LT of the expansion is equivalent to the result of the
intersection number computed within the \textit{relative} twisted cohomology
theory \cite{matsumoto2018relative}, obtained by means of the \textit{delta-forms}
introduced in \cite{Caron-Huot:2021xqj,Caron-Huot:2021iev}. Although derived
in the case of 1-form, the result of our analysis allows us to present an
simplification of the recursive algorithm for the evaluation of intersection
numbers of twisted $n$-forms, and to apply it to the complete decomposition of
a few non-trivial representative types of Feynman integrals at one and two
loops, with planar and non-planar configurations, in terms of bases of MIs.
On the one hand, we adapt the polynomial expansion technique introduced in
\cite{Fontana:2023amt}, by proposing a novel choice of the polynomial-ideal
generator in the intermediate layers of the recursive approach.
On the other hand, we eliminate the need for the analytic regulators, by
applying the \textit{relative} twisted cohomology theory
\cite{matsumoto2018relative,Caron-Huot:2021xqj,Caron-Huot:2021iev}, and
present a systematic algorithm for choosing multivariate
delta-forms as elements of the dual cohomology bases, to significantly simplify the computation of the intersection numbers.

The chosen examples of Feynman integrals involve differential $n$-forms with $n$ up to nine. The intersection-theory based decomposition is performed on cuts, along the lines of \cite{Frellesvig:2019kgj,Frellesvig:2020qot}, so that the determination of the coefficients of the integral decomposition require intersection numbers of $n$-forms, with $n$ up to six, only.
In particular, we describe in detail the decomposition via intersection numbers of integrals related to the one-loop box diagram contributing to Bhabha scattering, as well as the planar and non-planar massless double-box diagrams. Computational details for the latter two cases can be found in the ancillary files
\filename{Dbox\_massless.m} and
\filename{Dbox\_massless\_nonplanar.m}, respectively.
For the decomposition of the integrals related to the planar and non-planar
double-box diagrams with one massive external leg, in the text we provide just
the bases of $n$-forms and give the computational details, respectively, in the
ancillary files
\filename{Dbox\_1m.m}, and
\filename{Dbox\_1m\_nonplanar.m}. The decomposition formulas are found to be equivalent to the results of an IBP decomposition.
These results demonstrate that, using intersection numbers it is possible to obtain the {\it direct} and {\it complete} decomposition of non-trivial integrals in terms of MIs, just like any generic element of a vector space can be projected onto a set of generators.

This work is organized as follows:
In~\secref{sec:cohomologies} a review of the twisted de Rham cohomology and
intersection numbers is given.
\secref{sec:intersectionnumbers} contains a discussion of the application of polynomial
division and global residues for the efficient computation of intersection numbers.
\secref{sec:relative} describes the computation of intersection numbers within the relative
twisted cohomology theory using delta-forms, as well as their derivation as limits of cases with analytic regulators within the
regular twisted cohomology theory.
Applications to the decomposition of one- and two-loop Feynman integrals are
presented in~\secref{sec:examples} to showcase the novel techniques introduced
in this work.
We provide concluding remarks in \secref{sec:conclusions}.

The manuscript contains two appendices:
\appref{app:eea} contains a description of the extended euclidean algorithm
and \appref{sec:pickingbases} contains an example of our algorithm for building the bases for the (relative) twisted cohomology groups.

For our research, the following software has been used:
\soft{LiteRed}~\cite{Lee:2012cn, Lee:2013mka},
\soft{Fire}~\cite{Smirnov:2019qkx},
\soft{FiniteFlow}~\cite{Peraro:2019svx},
\soft{Fermat}~\cite{Fermat},
\soft{Fermatica}~\cite{Fermatica},
\soft{Mathematica},
\soft{HomotopyContinuation}~\cite{HomotopyContinuation.jl},
\soft{Singular}~\cite{DGPS} and its \soft{Mathematica} interface {\sc Singular.m}~\cite{SingularInterface}, and
\soft{Jaxodraw}~\cite{Binosi:2008ig, Axodraw:1994}.

\section{Integrals and Twisted Cohomology Groups}
\label{sec:cohomologies}
In this section we review some basic properties of twisted cohomology and
intersection theory, focusing on the evaluation of intersection numbers for
$1$- and $n$-forms.

The frameworks of \textit{twisted} homology and \textit{cohomology} are concerned with {\it twisted period integrals} of the form
\begin{align}
    I &= \int_{\mathcal{C}_R} \! u \ \phiL \defas \langle \phiL | \mathcal{C}_R ]
    \>,
    \label{eq:integraldef}
\end{align}
where $\phiL$ is a rational/meromorphic $n$-form:
$\phiL = {\widehat \varphi}_L(z) \, \dd \mathbf{z} \equiv {\widehat \varphi}_L(z) \,
\dd z_1 \wedge \ldots \wedge \dd z_n$,
and $u$ is a multivalued function
\begin{align}
    u = \prod_i B_i^{\gamma_i}
    \>,
    \label{eq:u_def}
\end{align}
called the \textit{twist}, with
generic \brk{polynomial} factors $B_i = B_i(z)$ and generic exponents
$\gamma_i\>$.
The generiticity condition on the $\gamma_i$ is required to ensure that
all poles of $\phiL$ are regulated by $u$.
Moreover, the integration domain ${\cal C}$ is chosen such that
$B_i(\partial {\cal C})=0$.
The latter condition ensures that
twisted period integrals obey integration-by-part identities (IBPs)
\begin{align}
    \int_{\mathcal{C}_R} \! \dd (u \, \phi_L) = 0 \>,
    \label{eq:def_ibps}
\end{align}
corresponding to the vanishing integrals
\begin{align}
\label{eq:ibpi}
\int_{\mathcal{C}_R} \! u \, \nabla_\omega \, \phi_L  =
\langle \nabla_\omega \, \phi_L | \mathcal{C}_R ] = 0 \>,
\end{align}
where we introduced the covariant derivative $\nabla_\omega\>$, defined as,
\begin{align}
    \nabla_\omega \defas \dd + \omega \>, \qquad
    \text{with} \;\;\qquad \omega \defas \dlog(u)
    = \sum_{i=1}^n \widehat{\omega}_i \, \dd z_i \>,
    \qquad \text{and} \;\;\qquad
    \widehat{\omega}_i \defas \partial_{z_i} \log(u) \>.
    \label{eq:omega_def}
\end{align}

Within twisted de Rham theory, any $n$-form $\phiL$
is a member (equivalence class representative) of the vector space of closed
modulo exact $n$-forms, called the $n$\textsuperscript{th} twisted cohomology
group $\rmH^{n}_\omega\>$. The equivalence relation reads $\langle \phiL | \sim \langle \phiL + \nabla_\omega \phi_L| $, where $\phi_L$ is a generic $(n{-}1)$-form. $\rmH^{n}_\omega$ identifies integrands that give the same
$I$, upon integration over ${\cal C}_R\>$.
The number $\nu \defas \Dim\lrbrk{\rmH_{\omega}^n}$ of
independent equivalence classes is the dimensionality of the
cohomology group.

We may also introduce the \textit{dual integrals}:
\begin{align}
    I^{\vee} &= \int_{\mathcal{C}_L} \! u^{-1} \ \phiR \defas [ \mathcal{C}_L | \phiR \rangle \>,
    \label{eq:dualintegraldef}
\end{align}
whose integrands contains the multivalued function $u^{-1}$ and the dual $n$-form
$\phiR\>$. Similarly to the previous case,
$| \phiR \rangle \sim
| \phiR + \nabla_{-\omega} \phi_R \rangle $ with $\phi_R$ being an $(n{-}1)$-form. Thus $\phiR$
is a member of another vector space of
closed modulo exact $n$-forms, called the $n$\textsuperscript{th} \textit{dual} twisted
cohomology group $\rmH^{n}_{-\omega}\>$. Elements of this group identify
integrands that give the same $I^\vee$, upon integration over ${\cal C}_L\>$.

Stokes' theorem motivates the examination of the equivalence classes of
integration domains ${\cal C}_{L,R}$ (called the $n$\textsuperscript{th}
twisted chains), which form the other two vector spaces $\rmH_{n}^{\pm \omega}$
referred to as the (dual) twisted $n$\textsuperscript{th} homology groups\footnote{
    We do not elaborate further on the homology groups, and refer the
    interested reader to \cite{hwa1966homology} for details.
}.
In turn, de Rham's theorem ensures that the four vector spaces
$\rmH^{n}_{\pm \omega}$ and $\rmH_{n}^{\pm \omega}$ are isomorphic,
which implies that
$\Dim\left( \rmH_\omega^{n} \right) = \Dim\left( \rmH_{-\omega}^{n} \right) = $
$\Dim\left( \rmH^\omega_{n} \right) = \Dim\left( \rmH^{-\omega}_{n} \right)$.

In the case of Feynman integrals, the dimension of the twisted cohomology group
corresponds to the number of master integrals:
\begin{align}
    \nu
    &= \Dim\left( \rmH^{n}_{\pm \omega} \right)
     = \Dim\left( \rmH_{n}^{\pm \omega} \right)\nonumber \\
    &= \text{number of zeros of $\omega$} \>,
    \label{eq:leepom}
\end{align}
which is determined by the number of critical points of the Morse height function $\log (\abs{u})$ \cite{Lee:2013hzt}. 

Let us consider the bases $\{ \langle e_i | \}_{i=1, \cdots ,\nu}$ belonging to $\rmH^{n}_{\omega}$ and the dual bases $\{ | h_i \rangle \}_{i=1, \cdots ,\nu}$ belonging to $\rmH^{n}_{-\omega}$. Then any arbitrary cocycle ($\langle \phiL |$) can be decomposed in terms of these bases following the \textit{master decomposition formula}~\cite{Mastrolia:2018uzb, Frellesvig:2019kgj}.
This relation expresses a given cocycle $\langle \phiL |$ in terms of a
given basis $\langle {e}_i |$:
\begin{align}
    \langle \phiL |
    =
    \sum_{i=1}^{\nu} c_i \langle {e}_i |
    \>,
    \qquad\quad
    \text{with}
    \qquad\quad
    c_i = \langle {\phiL} \,|\, {h}_j \rangle (\Cmat^{-1})_{ji}
    \>,
    \label{eq:masterdecomposition}
\end{align}
where the square matrix of all possible intersection numbers between the left-
and right-bases
\begin{align}
    \Cmat_{ij} \defas \langle {e}_i \,|\, {h}_j \rangle
    \label{eq:firstCdef}
\end{align}
is called the \textit{metric} or simply \textit{the $\Cmat$-matrix}.

Following the \textit{master decomposition formula}~\cite{Mastrolia:2018uzb, Frellesvig:2019kgj}, any Feynman integral $I$ can be decomposed in terms of the master integrals $J_{i}$ as
\begin{align}
    I = 
      \langle \phiL | \mathcal{C} ]\
    = \sum_{i=1}^{\nu} c_i \langle e_i | \mathcal{C} ]
    = \sum_{i=1}^{\nu} c_i \, J_i \, ,
\end{align}
where the coefficients of the decomposition $c_i$ are given by eq.~\eqref{eq:masterdecomposition} and 
$J_i = \langle e_i | \mathcal{C} \big]$ .
Let us observe that $c_i$ is independent of the choice of the dual bases $|h_j \rangle$ \cite{Frellesvig:2019kgj}, and that, this degree of freedom can be exploited to simplify the computing algorithm - as it soon will be made clear. 

\subsection{Intersection numbers for $1$-forms}
\label{subsec:evaluating_the_intersection_number}
The \textit{intersection number} is an integral of the product between the left and right forms. To define it consistently, one of the forms has to be
regulated by expressing it as a specific representative of its cohomology
class:
\begin{align}
    \langle \phiL \,|\, \phiR \rangle \defas
    \frac{1}{2 \pi i} \int_{\cX} \! \iota(\phiL) \wedge \phiR
    =
    -\frac{1}{2 \pi i} \int_{\cX} \! \phiL \wedge \iota(\phiR)
    \>,
    \label{eq:interexdefiota}
\end{align}
where the $\iota$-operator denotes the regularization procedure defined in the
univariate case as:
\begin{align}
    \iota(\phiL) \defas \phiL - \nabla_{\!  \omega} (h \psiL) \>, \qquad
    \iota(\phiR) \defas \phiR - \nabla_{\! -\omega} (h \psiR)
    \>,
\end{align}
with the Heaviside functions
\begin{align}
    h \defas \sum_{i \in \Poles_\omega} (1 - \theta_{z,i})
    \>,
    \qquad
    \theta_{z,i} \defas \theta(|z {-} z_i| - \epsilon)
    \>,
    \qquad
    \Poles_\omega \defas \bigbrc{\text{poles of $\omega$} }
    \>.
    \label{eq:hthetadefs}
\end{align}
The domain of integration in eq.~\eqref{eq:interexdefiota} is defined as
$\cX \defas \CP \setminus \Poles_\omega\>$, and the set of singularities
$\Poles_\omega$ includes also $z = \infty$.
The functions $\psiL$ and $\psiR$ are the solutions to the differential
equations:
\begin{align}
    \nabla_{\!  \omega} \psiL = \phiL \>, \qquad
    \nabla_{\! - \omega} \psiR = \phiR \>.
    \label{eq:psiunivariate}
\end{align}

To compute intersection numbers eq.~\eqref{eq:psiunivariate} must be solved
around each pole $p\in\Poles_\omega\>$. Considering the pole at $z=p$, the solution
around this pole formally reads \cite{matsumoto1998,matsumoto2018relative},
\begin{equation}\label{eq:matsumotocontour}
    \begin{aligned}
        \psiL_{,p}(z)&=\frac{1}{(\eta_{+}-1)\,u(z)}\int_{C_p(z)} \!\! u(t)\,\phiL(t)
        \>,
        \\
        \psiR_{,p}(z)&=\frac{u(z)}{(\eta_{-} - 1)} \int_{C_p(z)} \!\! \frac{\phiR(t)}{u(t)}
        \>.
    \end{aligned}
\end{equation}

\begin{figure}[H]
\centering
\includegraphicsbox{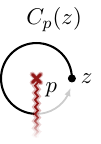}
    \caption{
        The integration contour $C_p(z)$ used in
        eq.~\protect\eqref{eq:matsumotocontour}.
    }
    \label{fig:contour1}
\end{figure}

Here $C_p(z)$ is the contour given in~\figref{fig:contour1} and $\eta_{\pm}=e^{\pm2\pi i
\alpha_p}$, with $\alpha_p$ being the non-integer exponent of $u$ at $z=p$
(and thus $-\alpha_p$ the exponent of $u^{-1}$).

Following \cite{Mastrolia:2018uzb, Frellesvig:2019kgj} we may then derive the expression for the univariate intersection number as
\begin{align}
    \vev{\phiL \,|\, \phiR} =
    \sum_{p \in \Poles_\omega} \Res_{z=p} ( \psiL_{,p} \, \phiR ) =
    -\sum_{p \in \Poles_\omega} \Res_{z=p} ( \psiR_{,p} \, \phiL )
    \>.
    \label{eq:univariate}
\end{align}

\subsection{Intersection numbers for $n$-forms}
\label{subsec:multivariate}
The intersection number for $n$-forms can be computed by adopting more than one computational strategy.
In this work, we will use the \textit{fibration}-based approach discussed in
\cite{Frellesvig:2019uqt, Frellesvig:2020qot}, 
which can be applied to generic meromorphic forms.
This method treats the integration variables one at a time,
so, without loss of generality, we may order them as
$z_n,\ldots,z_1$, listed from the outer to the innermost integration.
Each \textit{layer} of the fibration has its own (internal) basis of master forms,
whose size can be counted using eq.~\eqref{eq:leepom}, and can be efficiently chosen following the algorithm explained in~\secref{subsec:choice_of_basis_elements} and \appref{sec:pickingbases}.
We denote such bases of forms and their duals on the $m^\text{th}$
layer by $e\supbbf{m}_i$ and $h\supbbf{m}_i$ respectively, and their dimension
by $\nu_m$.

This approach allows us to compute multivariate intersection numbers recursively:
Assuming all the $\brk{m - 1}$-variate building blocks are known, the
$m$-variable intersection numbers can be evaluated using
\begin{align}
    \vev{
        \phiL^{\brkbf{m}} \,|\, \phiR^{\brkbf{m}}
    }
    =
    \sum_{p \in \Poles\supbrk{m}} \!\! \Res_{z_m=p} \left( \psiL_{,i}^{(m)}
    \Cmat^{(m-1)}_{ij} {\projPhi_{R,j}^{(m)}}
    \right)
    =
    \sum_{p \in \Poles\supbrk{m}} \!\! \Res_{z_m=p} \left( \psiL_{,i}^{(m)}
    \vev{e\supbbf{m - 1}_i \,|\, \phiR^{\brkbf{m}}}
    \right)
    \>,
    \label{eq:multivariate}
\end{align}
where the $\Cmat$-matrix and the projections are given by
\begin{align}
    \Cmat^{(m)}_{ij}
    &\defas
    \vev{
        e\supbbf{m}_i \,|\, h\supbbf{m}_j
    }
    \>,
    \\[1mm]
    \projPhi^{(m)}_{L,i}
    &=
    \sum_{j=1}^{\nu_{m-1}} \vev{
        \phiL^{\brkbf{m}} \,|\, h\supbbf{m - 1}_j
    }
    \,
    ( \Cmat_{(m-1)}^{-1} )_{ji}
    \>,
    \\
    \projPhi^{(m)}_{R,i}
    &=
    \sum_{j=1}^{\nu_{m-1}}
    ( \Cmat_{(m-1)}^{-1} )_{ij}
    \,
    \vev{
        e^{(\mathbf{m-1})}_j \,|\, \phiR^{\brkbf{m}}
    }
    \>.
    \label{eq:projections}
\end{align}

The key formula~\eqref{eq:multivariate} relies on the solution to the system of
differential equations:
\begin{align}
    \Bigsbrk{
        \delta_{ij}\partial_{z_m} + \Omega^{(m)}_{ji}
    }\,
    \psiL_{,j}^{(m)} = \projPhi_{L,i}^{(m)}
    \>,
    \quad\;\; \text{where} \quad\;\;
    \Omega^{(m)}_{ij}
    =
    \vev{
        \brk{
            \partial_{z_m} {+} \omega_m
        }
        e\supbbf{m - 1}_i
        \,|\,
        h\supbbf{m - 1}_k
    }
    \,
    \bigbrk{\Cmat_{(m-1)}^{-1}}_{kj} \,.
    \label{eq:psimultivariate}
\end{align}
The $\Omega\supbrk{m}$ matrix also known as the \textit{connection} matrix
defines the set of singularities $\Poles\supbrk{m}$ appearing in
eq.~\eqref{eq:multivariate} as
\begin{align}
    \Poles\supbrk{m} \defas \bigbrc{
        \text{poles of $\Omega\supbrk{m}$}
    }
    \>,
    \label{eq:def_poles}
\end{align}
which also includes $z = \infty$.
In complete analogy to the univariate case
of~\secref{subsec:evaluating_the_intersection_number}, we may write the dual
representation of eq.~\eqref{eq:multivariate} as
\begin{align}
    \vev{\phiL^{\brkbf{m}} \,|\, \phiR^{\brkbf{m}}}
    =
    - \hspace{-6pt}
    \sum_{p \in \Poles\supbrk{m}} \!\! \Res_{z_m=p} \Bigbrk{
        {\projPhi_{L,i}^{(m)}} \Cmat^{(m-1)}_{ij} \psiR_{,j}^{(m)}
    }
    =
    - \hspace{-6pt}
    \sum_{p \in \Poles\supbrk{m}} \!\! \Res_{z_m=p} \Bigbrk{
        \vev{
            \phiL^{\brkbf{m}}\vert h\supbbf{m - 1}_i
        }
        \,
        \psiR_{,i}^{(m)}
    }\,,
    \label{eq:multivariatefirstdual}
\end{align}
which makes use of the solution $\psiR$ to the dual differential equation
and the dual connection $\Omega^{\vee (m)}$ given by
\begin{align}
    \Bigsbrk{
        \delta_{ij}\partial_{z_m}+\Omega^{ \vee (m)}_{ij}
    }\,
    \psiR_{,j}^{(m)} = \projPhi_{R,i}^{(m)}
    \>,
    \quad \text{where} \quad\;
    \Omega^{\vee (m)}_{ij}
    =
    \bigbrk{\Cmat_{(m-1)}^{-1}}_{ik}
    \,
    \vev{
        e\supbbf{m - 1}_k
        \,|\,
        \brk{
            \partial_{z_m}{-}\omega_m
        } h\supbbf{m - 1}_j
    }
    \>.
    \label{eq:psimultivariatedual}
\end{align}

\section{Intersection Numbers and Polynomial Division}
\label{sec:intersectionnumbers}

In general, the evaluation of intersection numbers in
eqs.~\eqref{eq:interexdefiota, eq:multivariate} requires
the solution of the differential equations~\eqref{eq:psiunivariate, eq:psimultivariate},
and a residue operation. In the univariate case, these operations are
performed locally around each pole of the function $\omega$ defined
in eq.~\eqref{eq:omega_def}, and likewise with the poles of $\Omega$ in the multivariate case.
Individual contributions to the intersection
number from each pole may contain irrational terms, which only cancel upon
considering their sum. In this section we recall the idea proposed in
\cite{Fontana:2023amt} to show how intersection numbers can be computed bypassing the
precise identification of singularities of $\omega$, thus avoiding algebraic extensions:
polynomial division and global residues can be used to derive the solution of the differential
equation, and to extract the sum over local residues directly from
the remainders of iterated polynomial divisions.
In the application of these techniques, within the recursive algorithm for the
evaluation of intersection numbers, we propose the use of a novel polynomial,
built as the least common multiple of the denominators appearing in the
connection matrix.

\subsection{Polynomial decomposition and global residue}

\subsubsection*{Univariate polynomials}
The decomposition of a univariate polynomial $p(z)$ in terms of
another degree $\kappa$ polynomial $B(z)$ may be written as
\begin{eqnarray}
    p(z) & = & \sum_{i=0}^{max} p_i(z) \, B(z)^i
    \>,
    \quad \text{where} \quad
    p_i(z) = \sum_{j=0}^{\kappa-1} p_{ij} \, z^j
    \quad \text{and} \quad
    \kappa \> \defas \> \deg(B)
    \>.
    \label{eq:uni_pol_dec}
\end{eqnarray}
We observe that $p_i(z)$ correspond to the remainders of a sequential polynomial divisions of $p(z)$ and of the successive quotients w.r.t. $B(z)$ \cite{Fontana:2023amt}. By following the latter reference,
this operation can be efficiently obtained in one step by
introducing a shift parameter $\beta$ and a new divisor $\Baikov=\Baikov(z,\beta)$, defined as
\begin{align}
    \Baikov \defas B(z)-\beta
    \>,
    \label{eq:Baikov_def}
\end{align}
The remainder of a single division of $p(z)$ modulo $\Baikov$ belongs to the
quotient space ${\mathbb C}[z]/\langle \Baikov \rangle$ (where ${\mathbb C}[z]$
is the space of polynomials in the variable $z$, with complex coefficients, and
$\langle \Baikov \rangle$ is the ideal generated by $\Baikov(z, \beta)$),
is given by
\begin{eqnarray}
    \lfloor p(z) \rfloor_{\Baikov} \ =
    \sum_{i=0}^{max}\sum_{j=0}^{\kappa-1}p_{ij}\, z^j \, \beta^i
    \>.
\end{eqnarray}
We thus recover eq.~\eqref{eq:uni_pol_dec} upon
identifying $\beta = B(z)$.

\subsubsection*{Univariate rational functions}
Let us now consider a rational function $f(z)$, defined as the
ratio of two polynomials $n(z)$ and $d(z)$:
\begin{eqnarray}
    f(z) & = & \frac{n(z)}{d(z)}
    \>,
    \label{eq:f_func}
\end{eqnarray}
whose Laurent series expansion around $B(z)=0$ takes the form
\begin{eqnarray}
f(z) & = &
    \sum_{i=min}^{max}
    f_i(z)
    \, B(z)^i
    + \ \mathcal{O} \big( B(z)^{max+1} \big)
    \>,
    \quad {\rm with} \quad
    f_i(z) : = \sum_{j=0}^{\kappa-1} f_{ij} \, z^j
    \>.
    \label{eq:f_laurent}
\end{eqnarray}
Also in this case the polynomials $f_i(z)$ can be built as remainder of the
polynomial divisions of $f(z)$ modulo $\Baikov$, namely
\begin{eqnarray}
    \lfloor f(z)\rfloor_{\Baikov}  & = &
    \sum_{i=min}^{max}\sum_{j=0}^{\kappa-1} f_{ij} \, z^j
    \, \beta^i \;
    + \; \mathcal{O}(\beta^{max+1})
    \>,
    \label{eq:f_uni_exp}
\end{eqnarray}
and by identifying $\beta=B(z)$.
The rational function $f(z)$
is equivalent to the product of two polynomials $n(z)$ and
${\tilde d}(z)$:
\begin{eqnarray}
    \lfloor f\brk{z} \rfloor_{\Baikov}
    \ = \ \biggl\lfloor \frac{n(z)}{d(z)}\biggr\rfloor_{\Baikov}
    \ = \ \lfloor n(z) \, \tilde{d}(z)\rfloor_{\Baikov}
    \>.
\end{eqnarray}
Here, $\tilde{d}(z)$ is the
\textit{multiplicative inverse} of the denominator $d\brk{z}$ modulo $\Baikov$, defined as
\begin{eqnarray}
    \tilde{d}(z) \, d(z) \ = \ 1 \ \ mod \ \ \Baikov \>,
\label{eq:mult_inverse}
\end{eqnarray}
which can be determined%
\footnote{%
    The $\beta$-shift in definition~\eqref{eq:Baikov_def} ensures that
    eq.~\eqref{eq:mult_inverse} always has a solution
    in $\Complex\sbrk{z}/\vev{\Baikov}$.
}
either by ansatz, or, equivalently, by using the \textit{Extended Euclidean
Algorithm} \brk{see~\appref{app:eea} for details}.

\subsubsection*{Global residue}
Let us consider again the function $f(z)$ as in eq.~\eqref{eq:f_func}.  To
compute the global residue of $f\brk{z}$ over some polynomial $B(z)$,
which is the sum of the local residues evaluated at the zeroes of $B(z)$,
we may expand the function $f(z)$ around $\Baikov$ as in
eq.~\eqref{eq:f_uni_exp}, and then use the global residue theorem \brk{see~\cite{cattani} for review} to obtain:
\begin{eqnarray}
    \sum_{p\in \mathcal{P}_B} \! \Res_{z=p}(f(z))
    \; \safed \; \Res_{\langle B \rangle} (f(z) )\
    = \frac{f_{-1,\kappa-1}}{\ell_c} \>,
    \label{eq:global_residue}
\end{eqnarray}
where $\mathcal{P}_B\>$, $\ell_c\>$, and $\kappa$ are the set of zeroes, the
leading coefficient, and the degree of $B(z)$ respectively.

\subsection{Intersection numbers for $1$-forms and polynomial division}
The polynomial decomposition technique introduced in
the previous section can be applied to the computation of intersection numbers.
We consider first the univariate case.
To compute the global residue we choose the degree $\kappa$ polynomial
ideal generator
\begin{align}
    B\brk{z} \defas \LCM\bigbrk{\Poles_{\omega, \mathrm{fin}}}
    \label{eq:B_uni}
\end{align}
constructed via the \textit{least common multiple} $\LCM$
of the finite poles of $\omega$ introduced in
eq.~\eqref{eq:hthetadefs}.
The sum over the contributions to the intersection number~\eqref{eq:univariate}
stemming from the \textit{finite} poles
can be obtained as the global residue over the zeroes of $B$, namely
\begin{equation}
    \langle \phiL  \vert \phiR \rangle
    =
    - \Res_{\langle B \rangle} \bigbrk{g}
    \> - \>
    \Res_{z=\infty} \bigbrk{g}
    \>,
\end{equation}
where  $\psi_R$ satisfies \eqref{eq:psiunivariate}, and $g$ is defined as:
\begin{eqnarray}
    g\  = \ \psi_R\  \phiL
    \>.
    \label{eq:rho_uni}
\end{eqnarray}
The global residue can be computed via the polynomial division in the following
way:
\begin{enumerate}
    \item
    Compute the series expansions of $\phiL\>$, $\phiR\>$, and $\omega$ around $B(z) = 0$,
        given by $\lfloor\phiL\rfloor_{\Baikov}\>$,
        $\lfloor\phiR\rfloor_{\Baikov}\>$, and
        $\lfloor \omega\rfloor_{\Baikov}\>$ respectively,
        each having the form shown in eq.~\eqref{eq:f_uni_exp}.
    \item
     Build the ansatz
        \begin{eqnarray}
            \psi_R =
            \sum_{i=min}^{max}\sum_{j=0}^{\kappa-1} \psi_{R,ij} \, z^j
           \, \beta^i
         \>,
        \end{eqnarray}
        with unknown coefficients $\psi_{R,ij}\>$, and compute
        $\lfloor g \rfloor_{\Baikov}$ to extract $g_{-1,\kappa-1}$
        which depends on $\psi_{R,ij}\>$.
     \item
        Build the system of equations formed by the differential
        equation~\eqref{eq:psiunivariate} (using the composite derivative rule)
        together with the global residue~\eqref{eq:global_residue}:
        \begin{eqnarray}
            \biggl\lfloor
            \partial_z\psi_R(z,\beta) \
            + \ \partial_\beta \psi_R(z,\beta) \, \partial_z B(z)
            -  \omega \, \psi_R(z,\beta)
            - \phiR\ \biggr\rfloor_{\Baikov}
            &=& \ 0
            \>,
            \label{eq:DEQ_delta}
            \\
            \Res_{\langle B \rangle} (g) &=& \frac{g_{-1,\kappa-1}}{\ell_c}
            \>,
            \label{eq:global_res}
        \end{eqnarray}
        and solve it for $\Res_{\langle B \rangle} (g)$ only,
        obtaining the global residue directly from the linear system.
\end{enumerate}
Let us finally remark, that at all stages of the calculations $\beta$ can be
treated as a parameter (the actual substitutions $\beta \to B(z)$ is never
needed), which reduces the computational load of the problem.

\subsection{Intersection numbers for $n$-forms and polynomial division}
\label{ssec:n_poly}
The same technique can be applied in the multivariate case, following the
fibration approach described in \secref{subsec:multivariate}. Let us first
rewrite eq.~\eqref{eq:multivariatefirstdual} as:
\begin{eqnarray}
    \langle \phiL^{(\mathbf{m})} \vert \phiR^{(\mathbf{m})} \rangle
    & = &
    - \!\! \sum_{p\in \Poles\supbrk{m}_\mathrm{fin} \!} \!\!
    \Res_{z_m=p}\bigbrk{g^{(m)}}
    \> - \> \Res_{z_m=\infty}\bigbrk{g^{(m)}}
    \>,
    \label{eq:InterX_multi}
\end{eqnarray}
where $\Poles\supbrk{m}_\mathrm{fin}$ are the finite poles of $\Omega^{\vee (m)}$
\brk{cf. eq.~\eqref{eq:def_poles}}.
The function $g^{(m)}$ is defined as:
\begin{eqnarray}
    g^{(m)} \defas  \vev{\phiL\vert h_i^{(\mathbf{m}-1)}} \, \psi_{R,i}^{(m)} \>,
\end{eqnarray}
and $\psi_{R,i}^{(m)}$ satisfies eq.~\eqref{eq:psimultivariatedual}.
Similarly to the univariate case~\eqref{eq:B_uni}, to apply the
polynomial division algorithm at the $m$\textsuperscript{th} level of
iteration we use the polynomial ideal generator
\begin{eqnarray}
    B^{(m)}(z_m) \defas \LCM\bigbrk{\Poles\supbrk{m}_\mathrm{fin}}
    \label{eq:B_multi}
\end{eqnarray}
built out of the \textit{least common multiple} of the finite poles of
$\Omega^{\vee (m)}$ (namely out of the product of the denominators of its
entries, accounting for their highest multiplicity).
The regularised polynomial~\eqref{eq:Baikov_def} is then defined as
$\Baikov^{(m)}
\defas B^{(m)}(z_m) - \beta\>$, which allows us to rewrite the sum appearing in
eq.~\eqref{eq:InterX_multi} as a global residue:
\begin{eqnarray}
    \langle \phiL^{(\mathbf{m})} \vert \phiR^{(\mathbf{m})} \rangle
    & = &
    - \Res_{\langle B^{(m)} \rangle}\bigbrk{g^{(m)}}
    \> - \>
    \Res_{z_m=\infty}\bigbrk{g^{(m)}} \>.
\end{eqnarray}
The computation of the global residue can be carried out using polynomial division modulo the ideal
$\langle \Baikov^{(m)} \rangle$,
in full analogy with the univariate case, where, according to
eq.~\eqref{eq:global_residue}, the global residue is
given by:
\begin{eqnarray}
     \Res_{\langle B^{(m)} \rangle} (g^{(m)})= \frac{g^{(m)}_{-1,\kappa-1}}{\ell_c} \>,
\end{eqnarray}
with
\begin{eqnarray}
      \lfloor g^{(m)}\rfloor_{\Baikov^{(m)}} & = &
       \sum_{i=min}^{max}\sum_{j=0}^{\kappa-1} g^{(m)}_{ij} \, z_m^j \, \beta^i \
       + \ \mathcal{O}(\beta^{max+1}) \>,
\end{eqnarray}
where $\kappa$ and $\ell_c$ are the degree and the leading coefficient of
$B^{(m)}$ respectively.

Hence, intersection numbers for $n$-forms can be computed within the recursive
algorithm, using polynomial division and global residue at each step of the
sequence, by avoiding the calculation of the residue (and the solution of a
differential equation) around each pole, and keeping all the intermediate
calculations strictly rational, therefore allowing the use of finite fields
methods \cite{Peraro:2016wsq, Peraro:2019svx,Fontana:2023amt}.

\section{Relative Twisted Cohomology}
\label{sec:relative}
In this section, we extend our framework to relative twisted cohomology \cite{matsumoto2018relative}.
We recall the definition of delta-forms introduced in \cite{Caron-Huot:2021iev,Caron-Huot:2021xqj} 
 and show how, at least in the case of 1-forms, 
they emerge naturally when considering the series expansion of intersection numbers in limit of evanescent regulator parameter.

\subsection{Relative twisted cohomology and univariate delta-forms}
\label{subsec:relative_twisted_cohomology}
We might consider what happens if we relax the criterion of
eq.~\eqref{eq:integraldef}, requiring that all poles of $\phiL$ and $\phiR$ are
regulated by $u$.  In such a case, if the point $z=p$ is non regulated, a
local holomorphic solution $\psi_{p,L}$ of the differential equation~\eqref{eq:psiunivariate}
may not exist, therefore invalidating the algorithm
of Sections \ref{sec:cohomologies} and \ref{sec:intersectionnumbers}
for computing the intersection numbers.
\textit{Relative twisted cohomology}~\cite{matsumoto2018relative} offers the proper mathematical framework to
address such cases, where the contribution of the non-regulated poles to the
intersection numbers is efficiently evaluated through the use of $n$-forms
built with Dirac delta functions
\cite{matsumoto2018relative,Caron-Huot:2021xqj,Caron-Huot:2021iev}. These
forms play an essential role when used in the evaluation of the decomposition
coefficients eq.~\eqref{eq:masterdecomposition} where they are chosen as
elements of the dual bases. We will refer to them as \textit{delta-forms} in the rest of this work.

Let us first discuss the univariate case. If $z$ is unregulated
at point $z=0$ (which we pick without loss of generality), the
corresponding delta-form is defined as
\begin{align}
    \delta_{z} := \frac{u(z)}{u(0)} \, \dd \theta_{z,0}
    \>,
    \label{eq:deltadef}
\end{align}
where $\theta_{z,0}$ is defined in
eq.~\eqref{eq:hthetadefs}. That this is a valid right-form can be shown by the
fact that it is closed:
\begin{align}
    \nabla_{-\omega} \delta_z
    \>=\>
    \frac{\dd u}{u(0)} \> \dd \theta_{z,0}
    + \frac{u}{u(0)} \> \dd^2 \theta_{z,0}
    - \frac{\dd u}{u} \, \frac{u}{u(0)} \> \dd \theta_{z,0}
    \>=\>
    0
    \>.
\end{align}
For delta-forms, the $\iota$-regulation of eq.~\eqref{eq:interexdefiota} is
not needed and the intersection pairing can be defined directly.
In the univariate, case we may derive
\begin{align}
    \braket{ \phiL }{ \delta_z }
    := \frac{-1}{2 \pi i} \int_{\mathcal{X}} \! \phiL \wedge \delta_z
    = \Res_{z = 0} \left( \frac{u(z)}{u(0)} \phiL \right)
    \>,
    \label{eq:relativeunivariate}
\end{align}
in agreement with \cite{matsumoto2018relative, Caron-Huot:2021xqj,
Caron-Huot:2021iev}. We will discuss the multivariate analogue in \secref{subsec:n_rel}.

\subsection{From cohomology to relative cohomology}
By focusing our analysis to the case of 1-forms,
we show that the formula of the intersection number in ordinary twisted cohomology, when expressed as Laurent series for a vanishing regulator, contains the intersection numbers  
for relative twisted cohomology. 
This relation allow us to establish, an explicit, direct link between the results of~\cite{Frellesvig:2019kgj,Frellesvig:2019uqt} and
\cite{matsumoto2018relative,Caron-Huot:2021xqj,Caron-Huot:2021iev}, on the
one hand, and \cite{Fontana:2023amt}, on the other one.

\subsubsection*{$\psi$ in the vanishing regulator limit}
\label{subsec:potentialinrhoto0limit}
\begin{figure}[H]
\centering
\includegraphicsbox{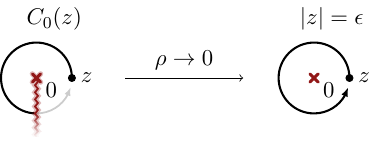}
    \caption{
        The integration contour $C_0(z)$
        used in eq.~\protect\eqref{eq:psi0}
        on the left and its leading
        contribution in the $\rho \to 0$ limit appearing in
        eq.~\protect\eqref{eq:contour_potential} on the right.
    }
    \label{fig:contour2}
\end{figure}

Let us consider the intersection number between two forms
$\braket*{\phiL}{\phiR}$ with a twist $u$, where $u$ does not have a branch
point at $z=0$.
If there exists the possibility that $\phiL$ or $\phiR$
have a pole at $z=0$ then, following \cite{Frellesvig:2019kgj,
Frellesvig:2019uqt}, a generic analytic regulator $\rho$ must be introduced,
modifying the twist to $u_\rho(z)=z^\rho\,u(z)$, such that $u_\rho$ is regulated around $z=0$. Using eq.~\eqref{eq:matsumotocontour} the
solution for $\psi$ at $z=0$ formally reads
\begin{equation}
    \psi_0(z)
    =
    \frac{z^\rho\,u(z)}{(e^{-2\pi i \rho}-1)}
    \int_{C_0(z)}t^{-\rho}\,
    \frac{\phiR(t)}{u(t)}
    \>.
    \label{eq:psi0}
\end{equation}
Let us now consider this solution in the limit $\rho\to0$. By series expanding around $\rho=0$ we obtain
\begin{equation}
    \begin{aligned}
        \frac{z^\rho\,u(z)}{(e^{-2\pi i \rho}-1)}
        & = -{1 \over 2 \pi i} \bigg( 
        {1 \over \rho} + \log(z) + i \pi
        + \mathcal{O} (\rho) \bigg) u(z)
        \\
        t^{-\rho}\,\frac{\phiR(t)}{u(t)}
        &=\left(1-\log(t)\rho + \mathcal{O}(\rho^2)\right)\frac{\phiR(t)}{u(t)} \>.
    \end{aligned}
\end{equation}
Combining these two expansions gives
\begin{equation}\label{eq:contour_potential}
    \begin{aligned}
        \psi_0(z)
        &=-\frac{u(z)}{2\pi i \rho}\int_{C_0(z)} \!\! \frac{\phiR(t)}{u(t)} \; + \; \mathcal{O}(\rho^0) \>,
        \\
        &=-\frac{u(z)}{2\pi i \rho}\oint_{\epsilon} \, \frac{\phiR(t)}{u(t)} \; + \; \mathcal{O}(\rho^0) \>,
    \end{aligned}
\end{equation}
where in the second line we used the fact that the leading order term of the
integrand is single valued, so the integration contour $C_0(z)$ reduces to a
small circle of radius $\epsilon$ encircling the origin,
see~\figref{fig:contour2}.
The function $\psi_0$ diverges in the limit $\rho \to 0$ for generic
$\varphi_{L/R}$ and has at most a simple pole in $\rho$.

\subsubsection*{Intersection numbers in the vanishing regulator limit}
\label{subsec:intnumbersinrho0limit}
For a regulated twist of the form $u_\rho=z^\rho\,u(z)$ we have concluded that
at most $\psi_0\sim 1/\rho$ for small $\rho$. By using an analogue of
eq.~\eqref{eq:contour_potential} for any $p\neq0\in\mathcal{P}_\varphi\>$, it is
not difficult to show that $\psi_p$ cannot have a term of the form
$\sim1/\rho$. Using eq.~\eqref{eq:univariate} we conclude that
\begin{equation}\label{eq:master_pole_formula}
\begin{aligned}
    \braket{\phiL}{\phiR}
    &=\frac{1}{2\pi i \rho}\Res_{z=0}\left(u(z)\,\phiL(z)\oint_{\epsilon} \, \frac{\phiR(t)}{u(t)}\right) \;+\; \mathcal{O}(\rho^0)
    \\[.2cm]
    &=\frac{1}{\rho}
    \Res_{z=0}\bigbrk{u(z)\,\phiL(z)}
    \times \Res_{z=0} \bigbrk{
        \phiR(z) / u(z)
    }
    \;+\; \mathcal{O}(\rho^0) \>.
\end{aligned}
\end{equation}
The term $\mathcal{O}(\rho^0)$ includes both higher term contributions in
$\rho$ coming from $\psi_0$ and all terms from the potentials $\psi_p$ with
$p\neq0$. 
It is important to note that if $\phiL$ or $\phiR$ do not have any pole
at $z=0$, then the intersection number is finite in the $\rho \to 0$ limit 
(because at least one of the two residues vanishes).

Given any two forms $\phiL$ and $\phiR$, such that $\phiR$ behaves as $\phiR \sim z^\tau$ around $z=0$, we define the {\it leading term} (LT) of the intersection number as
\begin{eqnarray}
\label{eq:LT}
\braket{\phiL}{\phiR}_{\rm LT} &:=&
\left\{
\begin{array}{c l}
\left.\braket{\phiL}{\phiR}\right\rvert_{\rho = 0} \>,
           \quad & \quad {\rm for} \quad \tau \ge 0 \>,
           \\[.2cm]
           \Res_{z=0}\bigbrk{u(z)\,\phiL(z)} 
    \times \Res_{z=0} \bigbrk{\phiR(z)/u(z)} \>,
           \quad & \quad {\rm for} \quad \tau < 0 \>. 
\end{array}
\right.
\end{eqnarray}
In the above formula, $\left.\braket{\phiL}{\phiR}\right\rvert_{\rho = 0}$ 
is equivalent to the intersection number $\braket{\phiL}{\phiR}$
evaluated using $u$ as twist, instead of $u_\rho$. 

\subsubsection*{Equivalence to delta-forms}
For the cases $\tau <0,$
the definition of the leading term in eq.~\eqref{eq:LT} can be related to the action
of the delta-forms introduced in~\secref{subsec:relative_twisted_cohomology}.
In particular, for a generic $\phiL\>$, and $\phiR=1/z$, we observe that the LT of intersection number in (ordinary) twisted cohomology reads as,
\begin{equation}
\label{eq:delta_basis_univariate_formula_from_limit}
    \begin{aligned}
        \braket*{\phiL}{\frac{1}{z}}_{\LT}
        &=\Res_{z=0}\bigg(\frac{u(z)}{u(0)}\,\phiL(z)\bigg) \> 
        = \braket{\phiL}{\delta_{z}} \ ,
    \end{aligned}
\end{equation}
where, on the rightmost side, we consider 
the result of eq.~\eqref{eq:relativeunivariate}, 
computed within relative twisted cohomology.
For cases where $\phiR$ has a pole of order $k$ this formula generalises to
\begin{equation}
\label{eq:delta_basis_univariate_formula_from_limit2}
    \braket*{\phiL}{\frac{1}{z^k}}_{\LT}=\Res_{z=0}\left(\frac{\partial_{z}^{(k-1)}u(z)}{(k-1)! \, u(0)}\phiL(z)\right)\,.
\end{equation}
In the language of delta-forms this is would be equivalent to considering dual basis elements as
\begin{equation}
    \delta_{z}^{(k)}\sim\frac{\partial_{k}^{(k-1)}u(z)}{u(0)}\,\dd \theta\,.
\end{equation}

\subsubsection*{Integral decompositions in the vanishing regulator limit}
\label{subsec:integraldecompositionsinrho0limit}
According to the master decomposition formula~\eqref{eq:masterdecomposition},
in presence of a regulator, the coefficients $c_i$ can be computed from intersection numbers, as:
\begin{eqnarray}\label{eq:IBP_decomposition}
    c_i = \lim_{\rho \to 0}
    \>
    \sum_{j=1}^\nu \braket*{\phiL}{h_j} \, {\bf C}^{-1}_{ji}
    \>.
\end{eqnarray}
with ${\bf C}_{ij} := \braket{ e_i }{ h_j }$. 
We can exploit the independence of $c_i$ on the dual bases $| h_j \rangle$ 
to simplify the above result. 
While computing the intersection numbers 
$\braket{\eta }{h_j}$, with $\eta \in \{\phiL, e_i\}$, 
if an element of the dual basis 
$|h_j\rangle$ 
behaves as $h_j \sim z^\tau$ with $\tau < 0$, around $z=0$, 
then it can be replaced by 
$\ket{h_j^\prime} :=$ $\ket{\rho\,h_j} = \rho \, \ket{h_j}$, 
without altering the final result of $c_i$.
Upon this substitution, according to eqs.~\eqref{eq:master_pole_formula, eq:LT},
the singular behaviour in $\rho$ is eliminated:
\begin{equation}
\braket{\eta}{h_j} \simeq {\braket{\eta}{h_j}_{\rm LT} \over \rho} 
\to 
\braket{\eta}{h_j^\prime} \simeq \braket{\eta}{h_j}_{\rm LT} \>
= 
\langle \eta \, | \, \delta_z^{(-\tau)} \rangle \ .
\end{equation}
Then, only the leading terms of the intersection numbers become relevant, so that $c_i$ reduces to\begin{eqnarray}
\label{eq:IBP_decomposition_LO}
    c_i
    = \sum_{j=1}^\nu\braket*{\phiL}{h_j}_{\LT} \,
    ({\bf C}_{\LT}^{-1})_{ji}\>,
\end{eqnarray}
where $({\bf C}_{\LT})_{ij}\defas\braket{e_i}{h_j}_{\LT}\>$.
This formula necessarily requires that ${\bf C}_{\LT}$ be {\it
invertible}: this condition is satisfied for bases chosen
according to the algorithm described in~\secref{subsec:choice_of_basis_elements} below.

Two observations are in order: 
for the evaluation of $\braket{\eta }{ h_j'}$, we use eqs.~(\ref{eq:delta_basis_univariate_formula_from_limit},\ref{eq:delta_basis_univariate_formula_from_limit2}), without needing to solve the 
differential equation~\eqref{eq:psiunivariate} around the pole $z=0$;
moreover, the idea of substituting $|h_j \rangle \to \rho |h_j \rangle$, borrowed from \cite{Fontana:2023amt}, is used here just formally in the derivation of eq.~(\ref{eq:IBP_decomposition_LO}), and it plays no explicit role in the actual evaluation algorithm.

Alternatively, as originally prescribed in \cite{Fontana:2023amt}, the simplified
formula~\eqref{eq:IBP_decomposition_LO}, can be efficiently evaluated by
extracting the leading term of the intersection numbers while
solving the (rightmost) differential equation~\eqref{eq:psiunivariate} 
around the non regulated pole, with the prescription 
$\ket{\phiR} \to \rho\, \ket{\phiR}\>$. Then, 
the solution $\psi$ of the differential equation, to be used in \eqref{eq:univariate}, 
is computed by ansatz, 
in the $\rho \to 0$ limit, holding the leading coefficients in $\rho$ only.

We consider our derivation of eq.~(\ref{eq:IBP_decomposition_LO}) within ordinary twisted cohomology 
one of the main results of this work, since it establishes a simple, clean link to relative cohomology \cite{Caron-Huot:2021xqj,Caron-Huot:2021iev}, and 
it provides a sound theoretical framework to the prescription of the choice of the dual bases suggested    in \cite{Fontana:2023amt}.

\subsection{Intersection numbers for $n$-forms in relative twisted cohomology}
\label{subsec:n_rel}
To avoid the use of regulators in the case of $n$-forms, 
we extend the outcome of the discussion on the 1-forms, and make use of 
multivariate delta-forms~\cite{Caron-Huot:2021iev,Caron-Huot:2021xqj}. 
If the variables $\{z_1,\ldots,z_m\}$ (out of a total of $n$) are non-regulated at the point $z_i=0$, the corresponding delta-form is defined as
\begin{align}
    \delta_{z_1,\ldots,z_m} := \frac{u}{u(0)} \bigwedge_{i=1}^m \dd \theta_{z_i,0}
    \>,
\end{align}
where we use the shorthand notation $u(0) \defas u|_{z_1 \rightarrow 0, \ldots, z_m \rightarrow 0}\>$.

In this case, the intersection pairing becomes
\begin{align}
    \vev{\phiL | \delta_{z_1,\ldots,z_m} \phi}
    \defas
    \frac{(-1)^n}{(2 \pi i)^n} \int_{\mathcal{X}_n} \!\! \phiL \wedge \iota(\delta_{z_1,\ldots,z_m} \phi)
    \>,
\end{align}
where the $\iota$-operator only regulates the integrations over the variables which are regulated by $u$, i.e. $z_{m+1}, \ldots, z_n\>$,
and $\phi = \widehat{\phi} \> \dd z_{m + 1} \wedge \ldots \wedge \dd z_n\>$.
From this we may derive
\begin{eqnarray}
    \vev{\phiL | \delta_{z_1,\ldots,z_m} \phi}
    &=&
    \frac{(-1)^{n-m}}{(2 \pi i)^{n-m}} \int_{\mathcal{X}_{m+1\ldots  n}}
    \hspace{-.5cm}
    \Res_{z_1=0,\ldots,z_m=0} \Bigbrk{ \tfrac{u}{u(0)} \phiL } \wedge \iota(\phi)
    \\
    &=&
    \bigvev{\ 
        \Res_{z_1=0,\ldots,z_m=0} \Bigbrk{
            \tfrac{u}{u(0)} \phiL
        }
        \> \big| \>
        \phi
    }
    \label{eq:relativemultivariate}
    \>,
\end{eqnarray}
where the $\brk{n{-}m}$-variate intersection number on the RHS should be computed as
in the ordinary (non-relative) by using eq.~\eqref{eq:multivariate}.
Let us remark that the ratio $u/u(0)$ in \eqref{eq:relativemultivariate} 
(resp. in eq.~\eqref{eq:relativeunivariate}, for the univariate case) 
is understood to be evaluated as a series expansion around $z_i=0$ for $i=1,\ldots,m$
(resp. around $z=0$, for the univariate case).
Therefore, if $\phiL$ has simple poles only, the $u/u(0)$-factor
has no effect in the evaluation of the residue, namely 
$\Res_{z=0} \big((u/u(0)) \, \phiL \big) = \Res_{z=0}(\phiL)$.
For this reason,
within our algorithm, when $\phiL$ has an higher order pole in the non-regulated
variables, we replace it by an equivalent element within the same cohomology group having just a simple pole, easily obtained by integration-by-parts 
\brk{see~\secref{sec:introductoryexample} for an example}.

\subsection{Choice of Basis Elements}
\label{subsec:choice_of_basis_elements}

In this section, we discuss an algorithm for the determination of the elements of the basis and dual basis.
First, we address the problem of providing a valid basis
of MIs for a given integral family. In other words, we are interested in
determining a basis, denoted by $e$, for the twisted cohomology group.

Let us consider for concreteness a problem depending on $\mathbf{z}=\{z_1,
\dots , z_{n}\}$ variables (within Baikov representation, the number of
integration variables amounts to the number of denominators and ISPs).\\
We identify a sector, denoted by $\mathcal{S}$, as a subset of the
integration variables, say for concreteness $\mathcal{S}=\{z_1, \dots , z_s \}
\subset \mathbf{z}$. A subsector is identified, in a natural way, as a subset
of $\mathcal{S}$.\\ Given $\mathcal{S}$, we can consider the corresponding
regulated twist, say $u_{\mathcal{S}}$ and the associated
$\omega_{\mathcal{S}}\>$, defined as
\begin{equation}
    u_{\mathcal{S}} = \prod_{z_i \in \mathcal{S}} z_i^{\rho_i} \cdot u
    \>,
    \qquad \omega_{\mathcal{S}} = \dlog u_{\mathcal{S}}
    \>.
\end{equation}
The number of zeros of $\omega_{\mathcal{S}}\>$, denoted by $\nu_{\mathcal{S}}$
(cf. eq.~\eqref{eq:leepom}), corresponds to the number of MIs in the sector
$\mathcal{S}$, including all its possible subsectors. Equivalently, it amounts
to the number of elements in the basis admitting at most $\{ z_1, \dots ,
z_s\}$ in the denominator (but not other $z_i\>$, with $z_i \notin
\mathcal{S}$).

Then, a possible strategy for the determination of the basis elements can be outlined as follows:\\
We order all the possible sectors, according to the number of their elements,
from the smallest to the largest and we create a list containing all the
elements in the basis. Clearly, in the input stage the list is empty; it is
updated according to the following steps
\begin{itemize}
    \item Consider a certain sector $\mathcal{S}$ and count the number of zeros
        of the corresponding $\omega_{\mathcal{S}}$ (cf.
        eq.~\eqref{eq:leepom}).
    \item Update the list of basis elements, without over-counting the elements
        already considered in all the possible subsectors (if any).\
    \item Iterate to the next sector in the list.
\end{itemize}
Once the list of all possible sectors is processed, the updated list will
contain all the basis elements. In the context of relative cohomology, a
dual basis$-$denoted by $h-$may be obtained by replacing inverse power of the
variables by the corresponding delta-form, i.e.:
\begin{align}
    h = e |_{(z_i z_j \cdots)^{-1} \, \rightarrow \ \delta_{z_i z_j \cdots}}
    \>.
\end{align}
An important comment is in order. While analyzing a certain sector, we may
encounter the situation in which more than one basis element has to be added
to the list. If this is the case, we are free to choose the new elements in
different ways (e.g. introducing ISPs or denominators raised to higher powers)
but we cannot guarantee that the new elements are independent. We may verify
the validity of our choice a posteriori, by checking that the corresponding $\Cmat$-matrix
is invertible, i.e. $\det \Cmat \neq 0$.\\
~\\
The strategy described above can also be applied in order to obtain a list of
basis elements for each layer in the fibration procedure. At any given layer,
the full set of variables is just a subset of the full set $\mathbf{z}$, and
the notion of sector has to be considered as a mathematical definition with no clear
physical analogue.\\
For an example of the use of this algorithm, see \appref{sec:pickingbases}.

\section{Applications}
\label{sec:examples}

In this section the usage of the polynomial division algorithm
(\secref{sec:intersectionnumbers}), and the relative twisted cohomology
framework (\secref{sec:relative}), are applied to efficiently decompose some
specific Feynman integrals. All examples are done using the \textit{bottom-up
decomposition}~\cite{Frellesvig:2020qot} in which the reduction is performed on
a \textit{spanning set of cuts}, defined as the minimal set of cuts (each
corresponding to a maximal cut of a sector) for which each master integral
appears at least once. On a given cut, integrals depend on fewer integration
variables and each cut is associated to a different twist.
Using this procedure, the reduction of the full 
integral (out of cuts) may be obtained by combining the results from the individual cuts.

We anticipate, that all the decomposition formulas obtained within the intersection-theory based approach, 
are verified to be equivalent to those obtained by means of IBPs.

\subsection{One-loop box for Bhabha scattering}
\label{sec:introductoryexample}
\begin{figure}[H]
    \centering
    \includegraphicsbox[]{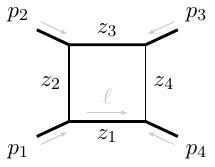}
    \caption{The one-loop box integral family contributing to Bhabha scattering.}
    \label{fig:bhabha}
\end{figure}

\begin{figure}[H]
    \centering
    \includegraphicsbox[scale=.6]{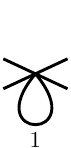}
    \>
    \includegraphicsbox[scale=.6]{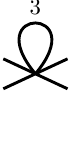}
    \>
    \includegraphicsbox[scale=.6]{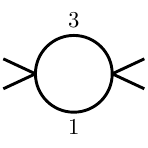}
    \>
    \includegraphicsbox[scale=.6]{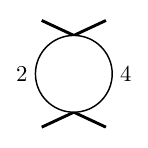}
    \>
    \includegraphicsbox[scale=.6]{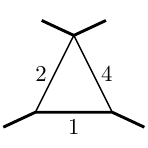}
    \>
    \includegraphicsbox[scale=.6]{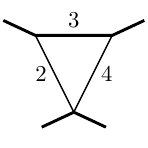}
    \>
    \includegraphicsbox[scale=.6]{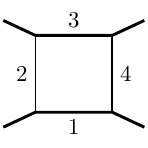}
    \caption{The $7$ master integrals for the one-loop box integral family contributing to Bhabha scattering, before imposing symmetry relations.} \label{fig:bhabha_MIs}
\end{figure}

As a warm-up example exhibiting the computational technology introduced above,
let us consider the one-loop box integral family contributing to Bhabha scattering shown in \figref{fig:bhabha}.
The denominators are chosen as
\begin{equation}
        z_1  =  \ell^2 -m^2, \quad
        z_2  =  (\ell-p_1)^2, \quad
        z_3  =  (\ell-p_1-p_2)^2-m^2, \quad
        z_4 =  (\ell -p_1-p_2-p_3)^2,
\end{equation}
while the kinematics is specified by
\begin{equation}
    p_i^2 \ = \ m^2 \ , \qquad s \ = \ (p_1 + p_2)^2 \ , \qquad t \ = \ (p_2+p_3)^2 \ , \qquad s  +  t  + u \ = \ 4 m^2 \>.
\end{equation}
There are $16$ possible sectors, and $7$ MIs depicted in \figref{fig:bhabha_MIs}
For illustration purposes we consider here the \textbf{Cut 24}, associated to $z_2=z_4=0$. On this cut just four MIs contribute and we focus for concreteness on the decomposition of the target integral:
\begin{equation}
    I  \ = \ \int \ \dd z_1 \dd z_3 \  u(z_1,z_3) \frac{1}{z_1z_3^2}
    \label{eq:box_tg}
\end{equation}
in terms of master integrals
\begin{equation}
    I \ = \ \sum_{i=1}^{4} c_i \ J_i
    \>,
\end{equation}
depicted as
\begin{equation}
    \vcenter{\hbox{\includegraphicsbox[scale=.6]{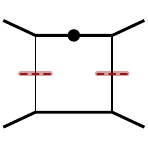}}}
    =
    c_1 \vcenter{\hbox{\includegraphics[scale=.6]{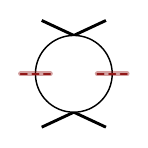}}}
    +
    c_2 \vcenter{\hbox{\includegraphics[scale=.6]{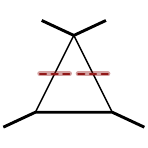}}}
    +
    c_3 \vcenter{\hbox{\includegraphics[scale=.6]{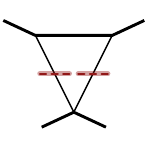}}}
    +
    c_4 \vcenter{\hbox{\includegraphics[scale=.6]{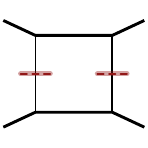}}}
    \>.
    \label{eq:boxdec}
\end{equation}
The twist is given by
\begin{eqnarray}
    u= B^ \gamma  \ , \qquad \text{where} \qquad \gamma=\frac{d-5}{2}
    \>,
\end{eqnarray}
with:
\begin{equation}
    B = -4 m^2 \left(s t+(z_1-z_3)^2\right)+s^2 t-2 s (t (z_1+z_3)+2 z_1 z_3)+t (z_1-z_3)^2.
\end{equation}
We may compute the connection as $\omega = \widehat{\omega}_1 \> \dd z_1 +
\widehat{\omega}_3 \> \dd z_3$ with
\begin{eqnarray}
    \widehat{\omega}_1 & = & \frac{(d-5) \left( s (t+2 z_3) + (4 m^2-t) (z_1-z_3) \right)}{4 m^2 \left(s t+(z_1-z_3)^2\right)-s^2 t+2 s t
   (z_1+z_3)+4 s z_1 z_3-t (z_1-z_3)^2}
    \>,
     \label{eq:omega1_bhabha_cut24}
    \\
   \widehat{\omega}_3 & = & \frac{(d-5) \left( s (t+2 z_1) - (4 m^2-t) (z_1-z_3) \right)}{4 m^2 \left(s t+(z_1-z_3)^2\right)-s^2 t+2 s t
   (z_1+z_3)+4 s z_1 z_3-t (z_1-z_3)^2}
    \>.
    \label{eq:omega3_bhabha_cut24}
\end{eqnarray}
Choosing the variable order $\{z_1,z_3\}$ and then following the procedure
of \appref{sec:pickingbases}, we get as the dimension of the inner and outer bases
\begin{equation}
    \nu\layer{3} \ = \ 2
    \>, \qquad
    \nu\layer{13} \  = \ 4
    \>.
\label{eq:counting_bhabha_cut24}
\end{equation}
The bases are
\begin{equation}
\begin{split}
    e\layer{3}  = \left\{ 1 \, , \; \frac{1}{z_3} \right\}, \qquad
    e=e\layer{13}  = \left\{ 1 \, , \; \frac{1}{z_1} \, , \; \frac{1}{z_3}  \, , \; \frac{1}{z_1 z_3} \right\}
    \>,
\end{split}
\end{equation}
while the dual bases are chosen as
\begin{equation}
\begin{split}
    h\layer{3}  = \{ 1 , \delta_3 \}, \qquad
    h =h\layer{13} = \{ 1 , \delta_1 , \delta_3 , \delta_{13} \}
    \>.
\end{split}
\end{equation}
The target left form associated to the integral on the LHS of eq.~\eqref{eq:boxdec}
is $\phiL = \tfrac{1}{z_1 z_3^2}\>$.

\subsubsection*{Computation}
\paragraph{Inner layer $z_3$:}
We first have to compute the $\Cmat$-matrix for the inner layer.

In order to do so we first find the
$\Baikov \ = \ B^{(3)}-\beta$
function for the internal layer, which is needed to compute the
sum over finite poles using the polynomial residue algorithm explained in
\secref{sec:intersectionnumbers}. $B^{(3)}$ is the denominator of $\Omega^{\vee
(3)}=-\omega_3\>$, which in this case corresponds to the Baikov polynomial, given
by:
\begin{eqnarray}
     B^{(3)} \ = \ 4 m^2 \left(s t+(z_1-z_3)^2\right)-s^2 t+2 s t (z_1+z_3)+4 s z_1 z_3-t (z_1-z_3)^2
     \>.
\end{eqnarray}
We notice that $z_3 = 0$ is not a zero of $B^{(3)}$, since it is an unregulated pole.
The intersection numbers between cocycles which do not contain any delta-forms are
evaluated as in eq.~\eqref{eq:univariate}
\begin{eqnarray}
    \langle e_i\layer{3} \vert h_1\layer{3} \rangle & = &  -\text{Res}_{\langle B^{(3)} \rangle} (e_i\layer{3} \psi_R^{(3)} ) \ - \ \text{Res}_{z =  \infty} ( e_i\layer{3} \psi_R^{(3)})
    \>,
\end{eqnarray}
while intersections with delta-forms read
\begin{eqnarray}
    \langle e_i\layer{3} \vert h_2\layer{3} \rangle &  = & \langle e_i\layer{3} \vert \delta_{3} \rangle \ = \ \text{Res}_{z_3=0} ( e_i\layer{3} )
    \>.
\end{eqnarray}
Performing the computations one gets the internal $\Cmat$-matrix, which is then given by:
\begin{eqnarray}
    \Cmat^{(3)} & = & \left(
\begin{array}{cc}
 \frac{- 4 (d-5) s \left(4 m^2 - s - t\right) \left(m^2 t+z (t+z)\right)}{(d-6) (d-4) \left(4 m^2 - t\right)^2} \; & 0 \\
 \frac{z \left(4 m^2 - 2 s - t\right) - s t}{(d-6) \left(4 m^2-t\right)} & 1
\end{array}
\right)
\>.
\end{eqnarray}

\paragraph{Outer layer $z_1$:}
The $\Omega^{\vee (1)}$ matrix for the outer layer is given as in eq.~\eqref{eq:psimultivariatedual}:
\begin{eqnarray}
   \Omega^{\vee (1)} \ = \ \left(
\begin{array}{cc}
 \frac{(d-6) (t+2 z_1)}{2 \left(m^2 t+z_1 (t+z_1)\right)} & 0 \\
 \frac{t \left(2 m^2-s+z_1\right)}{2 \left(m^2 t+z_1 (t+z_1)\right)} & \; \frac{(d-5) \left(4 m^2 z_1+s t-t
   z_1\right)}{4 m^2 \left(s t+z_1^2\right) - t (s-z_1)^2} \\
\end{array}
\right)
\>.
\end{eqnarray}

In order to use the polynomial division algorithm in the outer layer, one has
to find the corresponding ideal generator $\Baikov\layer{1}_{\beta} \ = \  B\layer{1}-\beta$
needed to compute the sum over finite poles. According to eq.~\eqref{eq:B_multi}, it
is given in terms of
\begin{eqnarray}
    B\layer{1} & =  & \LCM
    ( \mathcal{P}\layer{1}_{\mathrm{fin}} ) \ = \  2 \left(m^2 t+z_1 (t+z_1)\right) \left(4 m^2 \left(s t+z_1^2\right)-t (s-z_1)^2\right)
    \>.
\end{eqnarray}
The $\Cmat$-matrix for the outer layer reads:
\begin{equation}
    \Cmat^{(1)} \ = \ \left(
\begin{array}{cccc}
 \frac{s t^2 \left(4 m^2-s-t\right)}{4 (d-7) (d-3)} & 0 & 0 & 0 \\
 \frac{s t^2 \left((8 d - 44) m^2 - (d-6) t\right) \left(4 m^2-s-t\right)}{2 (d-7) (d-6) (d-4) \left(4 m^2-t\right)^2} & -\frac{4 (d-5) m^2
   s t \left(4 m^2-s-t\right)}{(d-6) (d-4) \left(4 m^2-t\right)^2} & 0 & 0 \\
 \frac{s t^2 \left((8 d - 44) m^2 - (d-6) t\right) \left(4 m^2-s-t\right)}{2 (d-7) (d-6) (d-4) \left(4 m^2-t\right)^2} & 0 & -\frac{4 (d-5)
   m^2 s t \left(4 m^2-s-t\right)}{(d-6) (d-4) \left(4 m^2-t\right)^2} \; & 0 \\
 \frac{s t^2}{(d-7) (d-6) \left(4 m^2-t\right)} & \frac{- s t}{(d-6) \left(4 m^2-t\right)} & \frac{-s t}{(d-6) \left(4 m^2-t\right)} & 1 \\
\end{array} \right)
\>.
\end{equation}

When computing the intersection numbers of ${\varphi}$ in the outer basis, we remove the higher order pole as:
\begin{eqnarray}
    {\varphi} & =  & \frac{1}{z_1 z_3^2} \ \sim \ \frac{\partial_{z_3}(u)}{u} \frac{1}{z_1 z_3} \ = \ \frac{\omega_3}{z_1 z_3} \ ,
    \label{eq:ex_ibp_trick}
\end{eqnarray}
where the $\sim$ means that the forms belong to the same cohomology class.
Then we get:
\begin{eqnarray}
    \langle{\varphi}\vert h_j\layer{13} \rangle & =  &\left(
\begin{array}{cccc}
 \frac{(d-5) s t}{(d-7) (d-6) \left(4 m^2-t\right)}\  , \ \  &  \frac{d-5}{6-d}\ ,  &\ \ \frac{(d-5) \left(4 m^2-2 s-t\right)}{(d-6) \left(4 m^2-t\right)}\ ,  & \ \
   \frac{d-5}{4 m^2-s} \\
\end{array}
\right) \>.
\end{eqnarray}
Combining this result with the metric $\Cmat^{(1)}$ via the master decomposition formula eq.~\eqref{eq:masterdecomposition} we get the coefficients of the reduction:
\begin{eqnarray}
    \langle {\varphi} \vert \mathcal{C}] & = & c_1 \langle e_1 \vert \mathcal{C}]\ + \ c_2 \langle e_2 \vert \mathcal{C}]\ +\  c_3 \langle e_3 \vert \mathcal{C}]  \ + \ c_4 \langle e_4 \vert \mathcal{C}] \ ,
\end{eqnarray}
which are given by:
\begin{eqnarray}
& & c_1 \ = \  -\frac{d-3}{m^2 t \left(4 m^2-s\right)} \ , \qquad\qquad \ c_2 \ =  \ \frac{(d-4) \left(4 m^2-t\right)}{s t \left(4 m^2-s\right)} \ , \nonumber \\
& &  c_3 \ = \ -\frac{(d-4) \left(2 m^2-s\right) \left(4 m^2-t\right)}{2 m^2 s t \left(4 m^2-s\right)} \ , \qquad \  c_4 \ = \ \frac{d-5}{4 m^2-s} \ ,
\end{eqnarray}
and are in agreement with FIRE.

More generally, the full decomposition can be achieved by combining the decompositions on different cuts, and in the
following we will see it in some two-loop examples.

\subsection{Planar double-box}
\label{subsec:planardb}

\begin{figure}
\centering
\includegraphics[width=0.9\textwidth]{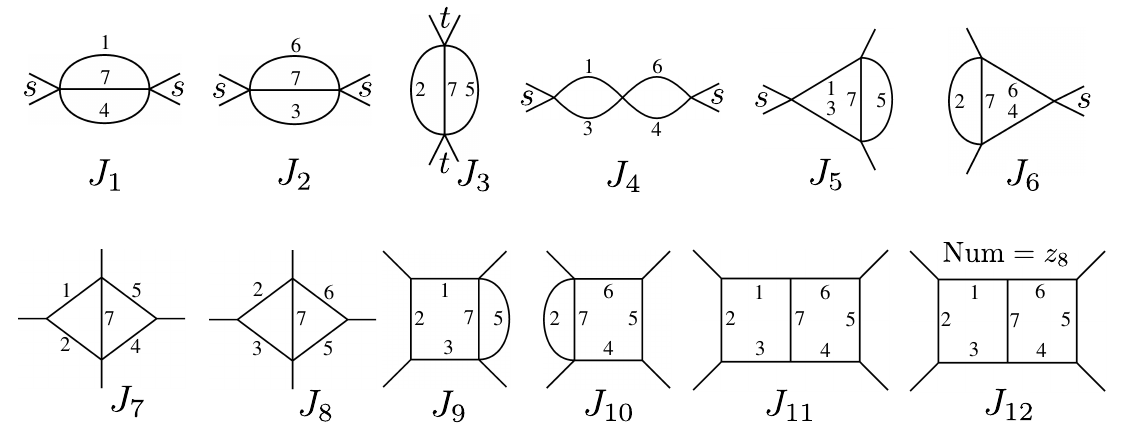}
\caption{The 12 master integrals for the planar double box integral family, before imposing symmetry relations.
    The index $i$, next to the propagators, indicates the corresponding $z_i$ variable;
    $\mathrm{Num}$ stands for the numerator factor; $s, t,$ and $u$ channels are indicated, to distinguish among graphs with identical shape, but corresponding to different integrals.
    \label{fig:doubleboxmasters}
}
\end{figure}

The integral family of the planar double-box is given in terms of
\begin{align}
z_1 &= k_1^2\>, & z_2 &= (k_1{-}p_1)^2, & z_3 &= (k_1{-}p_1{-}p_2)^2, & z_4 &= (k_2{-}p_1{-}p_2)^2, & z_5 &= (k_2{+}p_4)^2, \nonumber \\
z_6 &= k_2^2\>, & z_7 &= (k_1{-}k_2)^2, & z_8 &= (k_1{+}p_4)^2, & z_9 &= (k_2{-}p_1)^2, &
\end{align}
where $z_8$ and $z_9$ are irreducible scalar products, hence they may only appear in the numerator. The kinematics is such that:
\begin{equation}
    p_i^2 \ = \ 0 \;, \qquad s \ = \ (p_1{+}p_2)^2 \;, \qquad t \ = \ (p_1{+}p_4)^2 \;, \qquad s+t+u \ = \ 0 \;.
\end{equation}
This integral family has (before application of the symmetry relations) 12
master integrals, which we may pick as depicted in \figref{fig:doubleboxmasters}.
We are interested in decomposing the target integral:
\begin{equation}
    I  \ = \ \int \ \dd\mathbf{z}\  u(\mathbf{z}) \frac{z_8^2}{z_1z_2z_3z_4z_5z_6z_7}
    \label{eq:dbox_I}
\end{equation}
in terms of master integrals via a complete set of spanning cuts, as:
\begin{equation}
    I \ = \ \sum_{i=1}^{12} c_i \ J_i\>.
\end{equation}
The explicit expressions for the twist $u\brk{\mathbf{z}}$ as well as the master
integrals $J_i$ can be found in the ancillary
file~\filename{Dbox\_massless.m}.

The set of spanning cuts is given by the maximal cuts
of the first six master integrals $\brc{J_1, \ldots, J_6}$, and we will now go
through them one by one.

\subsubsection*{Cut 147, maximal cut of $J_1$}
Setting $z_1\ =\ z_4 \ = \ z_7 \ = \ 0$, and choosing as order of variables, from outer to inner:
\begin{equation}
    \{ z_3 , \ z_8 , \ z_2 , \ z_6 , \ z_5,\ z_9 \}
\end{equation}
we get as dimensions for the various layers:
\begin{equation}
    \nu\layer{9} \ = \ 1
    \>, \quad
    \nu\layer{59} \ = \ 2
    \>, \quad
    \nu\layer{659} \ = \ 2
    \>, \quad
    \nu\layer{2659}\ = \ 4
    \>, \quad
    \nu\layer{82659} \ = \ 5
    \>, \quad
    \nu\layer{382659} \ = \ 4
    \>.
\end{equation}
We pick as bases:
\begin{align}
&\qquad\qquad
\begin{aligned}
    e^{(9)} & =  \left\{1\right\}
    \>,
    &
    e^{(59)} & = \left\{1,\tfrac{1}{z_5}\right\}
    \>,
    &
    e^{(659)}& =  \left\{1,\tfrac{1}{z_5z_6}\right\}
    \>,
    \nonumber
\end{aligned}
\\
&
\begin{aligned}
    e^{(2659)} & = \left\{1,\tfrac{1}{z_2},\tfrac{1}{z_5 z_6},\tfrac{1}{z_2 z_5 z_6}\right\}
    \>,
    &
    e^{(82659)}&=\left\{1 , \tfrac{1}{z_5}  ,  \tfrac{1}{z_2 z_5} ,  \tfrac{1}{z_2 z_5 z_6}  ,  \tfrac{z_8}{z_2 z_5 z_6} ,\right\}
    \>.
    \nonumber
\end{aligned}
\\
&\qquad\qquad\qquad
\begin{aligned}
    e&=e^{(382659)} = \left\{1  ,   \tfrac{1}{z_2 z_5}  , \tfrac{1}{z_2 z_3 z_5 z_6},\tfrac{z_8}{z_2 z_3 z_5 z_6}\right\}
\>.
\end{aligned}
\end{align}
and as corresponding dual ones:
\begin{align}
&\qquad\qquad
\begin{aligned}
    h^{(9)} & = \left\{1\right\} \>, &
    h^{(59)} & = \left\{1,\delta_{5}\right\} \>, &
    h^{(659)} & = \left\{1,\delta_{56}\right\} \>,
    \nonumber
\end{aligned}
\\
&
\begin{aligned}
    h^{(2659)} & = \left\{1,\delta_{2},\delta_{56},\delta_{256}\right\} \>, &
    h^{(82659)} & = \left\{1,\delta_{5},\delta_{25},\delta_{256},z_8\delta_{256}\right\} \>,
    \nonumber
\end{aligned}
\\
&\qquad\qquad\,\,
\begin{aligned}
    h & = h^{(382659)} = \left\{1,\delta_{25},\delta_{2356},z_8\delta_{2356}\right\} \>.
\end{aligned}
\end{align}
We can then decompose the target left form: $\varphi \ =\ \frac{z_8^2}{z_2 z_3 z_5 z_6}$ in terms of the outer basis, where we omit the superscripts for simplicity, obtaining:
\begin{equation}
    \langle \varphi \vert \mathcal{C}] \ = \ c_1 \langle e_1 \vert \mathcal{C}] \ + \  c_7 \langle e_2 \vert \mathcal{C}] \ + \  c_{11} \langle e_3 \vert \mathcal{C}] \  + \  c_{12} \langle e_4 \vert \mathcal{C}]
    \>.
\end{equation}

\subsubsection*{Cut 367, maximal cut of $J_2$}
Setting $z_3\ =\ z_6 \ = \ z_7 \ = \ 0$, and choosing as order of variables, from outer to inner:
\begin{equation}
    \{ z_1 , \ z_8 , \ z_2 , \ z_5 , \ z_4,\ z_9 \}
\end{equation}
we get as dimensions for the various layers:
\begin{equation}
    \nu\layer{9} \ = \ 1
    \>, \quad
    \nu\layer{49} \ = \ 2
    \>, \quad
    \nu\layer{549} \ = \ 2
    \>, \quad
    \nu\layer{2549}\ = \ 4
    \>, \quad
    \nu\layer{82549} \ = \ 5
    \>, \quad
    \nu\layer{182549} \ = \ 4
    \>.
\end{equation}
We pick as bases:
\begin{align}
&\qquad\qquad\qquad
\begin{aligned}
    e^{(9)} & = \left\{1\right\} \>,
    &
    e^{(49)} & = \left\{1,\tfrac{1}{z_4}\right\} \>,
    &
    e^{(549)} & = \left\{1,\tfrac{1}{z_4z_5}\right\} \>,
    \nonumber
\end{aligned}
\\
&
\begin{aligned}
    e^{(2549)} & = \left\{1,\tfrac{1}{z_2},\tfrac{1}{z_4 z_5},\tfrac{1}{z_2 z_4 z_5}\right\} \>,
    &
    e^{(82549)} & = \left\{1,\tfrac{1}{z_5},\tfrac{1}{z_2 z_5},\tfrac{1}{z_2 z_4 z_5},\tfrac{z_8}{z_2 z_4 z_5}\right\} \>,
    \nonumber
\end{aligned}
\\
&\qquad\qquad\qquad\,\,
\begin{aligned}
    e & = e^{(182549)} = \left\{1,\tfrac{1}{z_2 z_5},\tfrac{1}{z_1 z_2 z_4 z_5},\tfrac{z_8}{z_1 z_2 z_4 z_5}\right\} \>.
\end{aligned}
\end{align}
and as corresponding dual ones:
\begin{align}
&\qquad\qquad
\begin{aligned}
    h^{(9)} & = \left\{1\right\} \>, &
    h^{(49)} & = \left\{1,\delta_{4}\right\} \>, &
    h^{(549)} & = \left\{1,\delta_{45}\right\} \>,
    \nonumber
\end{aligned}
\\
&
\begin{aligned}
    h^{(2549)} & = \left\{1,\delta_{2},\delta_{45},\delta_{245}\right\} \>, &
    h^{(82549)} & = \left\{1,\delta_{5},\delta_{25},\delta_{245},z_8\delta_{245}\right\} \>,
    \nonumber
\end{aligned}
\\
&\qquad\qquad\,\,
\begin{aligned}
    h & = h^{(182549)} = \left\{1,\delta_{25},\delta_{1245},z_8\delta_{1245}\right\} \>.
\end{aligned}
\end{align}
We can then decompose the target left form: $\varphi \ =\ \frac{z_8^2}{z_1 z_2 z_4 z_5}$ in terms of the outer basis, where we omit the superscripts for simplicity, obtaining:
\begin{equation}
    \langle \varphi \vert \mathcal{C}] \ = \ c_2 \langle e_1 \vert \mathcal{C}] \ + \  c_8 \langle e_2 \vert \mathcal{C}] \ + \  c_{11} \langle e_3 \vert \mathcal{C}] \  + \  c_{12} \langle e_4 \vert \mathcal{C}]
    \>.
\end{equation}

\subsubsection*{Cut 257, maximal cut of $J_3$}
Setting $z_2\ =\ z_5 \ = \ z_7 \ = \ 0$, and choosing as order of variables, from outer to inner:
\begin{equation}
    \{ z_1 , \ z_8 , \ z_3 , \ z_6 , \ z_4,\ z_9 \}
\end{equation}
we get as dimensions for the various layers:
\begin{equation}
    \nu\layer{9} \ = \ 1
    \>, \quad
    \nu\layer{49} \ = \ 2
    \>, \quad
    \nu\layer{649} \ = \ 2
    \>, \quad
    \nu\layer{3649}\ = \ 6
    \>, \quad
    \nu\layer{83649} \ = \ 10
    \>, \quad
    \nu\layer{183649} \ = \ 7
    \>.
\end{equation}
We pick as bases:
\begin{align}
&\qquad\qquad
\begin{aligned}
    e^{(9)} & = \left\{1\right\} \>, &
    e^{(49)} & = \left\{1,\tfrac{1}{z_4}\right\} \>, &
    e^{(649)} & = \left\{1,\tfrac{1}{z_4z_6}\right\} \>,
    \nonumber
\end{aligned}
\\
&\qquad\qquad\qquad\,\,\,\,\,
\begin{aligned}
    e^{(3649)} & = \left\{1,\tfrac{1}{z_3},\tfrac{1}{z_4},\tfrac{1}{z_4 z_6},\tfrac{z_9}{z_4 z_6},\tfrac{1}{z_3 z_4 z_6} \right\} \>,    
    \nonumber
\end{aligned}
\\
&\qquad\,\,
\begin{aligned}
    e^{(83649)} & = \left\{1,\tfrac{1}{z_3},\tfrac{1}{z_4},\tfrac{z_8}{z_4},\tfrac{1}{z_6},\tfrac{1}{z_3 z_6},\tfrac{1}{z_4z_6},\tfrac{z_8}{z_4 z_6},\tfrac{1}{z_3 z_4 z_6},\tfrac{z_8}{z_3 z_4 z_6}\right\} \>,    
    \nonumber
\end{aligned}
\\
&\qquad\,\,
\begin{aligned}
    e & = e^{(183649)} = \left\{1,\tfrac{1}{z_1 z_4},\tfrac{1}{z_3 z_6},\tfrac{1}{z_1 z_3},\tfrac{1}{z_4 z_6},\tfrac{1}{z_1 z_3 z_4 z_6},\tfrac{z_8}{z_1 z_3 z_4 z_6}\right\} \>.
\end{aligned}
\end{align}
and as corresponding dual ones:
\begin{align}
&\qquad\qquad
\begin{aligned}
    h^{(9)} & = \left\{1\right\} \>, &
    h^{(49)} & = \left\{1,\delta_{4}\right\} \>, &
    h^{(649)} & = \left\{1,\delta_{46}\right\} \>,
    \nonumber
\end{aligned}
\\
&\qquad\qquad\qquad
\begin{aligned}
    h^{(3649)} & = \left\{1,\delta_{3},\delta_{4},\delta_{46},z_9\delta_{46},\delta_{346}\right\} \>,    
    \nonumber
\end{aligned}
\\
&\qquad\,\,\,
\begin{aligned}
    h^{(83649)} & = \left\{1,\delta_{3},\delta_{4},z_8\delta_{4},\delta_{6},\delta_{36},\delta_{46},z_8\delta_{46},\delta_{346},z_8\delta_{346}\right\} \>,    
    \nonumber
\end{aligned}
\\
&\qquad\,\,\,
\begin{aligned}
    h & = h^{(183649)} = \left\{1,\delta_{14},\delta_{36},\delta_{13},\delta_{46},\delta_{1346},z_8\delta_{1346}\right\} \>.
\end{aligned}
\end{align}
We can then decompose the target left form: $\varphi \ =\ \frac{z_8^2}{z_1 z_3 z_4 z_6}$ in terms of the outer basis, where we omit the superscripts for simplicity, obtaining:
\begin{equation}
    \langle \varphi \vert \mathcal{C}] \ = \ c_3 \langle e_1 \vert \mathcal{C}] \ + \  c_7 \langle e_2 \vert \mathcal{C}] \ + \  c_{8} \langle e_3 \vert \mathcal{C}] \  + \  c_{9} \langle e_4 \vert \mathcal{C}] \ + \ c_{10} \langle e_5 \vert \mathcal{C}] \ + \ c_{11} \langle e_6 \vert \mathcal{C}]  \ + \ c_{12} \langle e_7 \vert \mathcal{C}]
    \>.
\end{equation}

\subsubsection*{Cut 1346, maximal cut of $J_4$}
Setting $z_1\ =\ z_3 \ = \ z_4 \ = \ z_6 \ = \ 0$, and choosing as order of variables, from outer to inner:
\begin{equation}
    \{ z_8 , \ z_2 , \ z_5 , \ z_7 , \ z_9 \}
\end{equation}
we get as dimensions for the various layers:
\begin{equation}
    \nu\layer{9} \ = \ 1
    \>, \quad
    \nu\layer{79} \ = \ 2
    \>, \quad
    \nu\layer{579}\ = \ 2
    \>, \quad
    \nu\layer{2579} \ = \ 4
    \>, \quad
    \nu\layer{82579} \ = \ 3
    \>.
\end{equation}
We pick as bases:
\begin{align}
&\qquad\qquad
\begin{aligned}
    e^{(9)} & = \left\{1\right\}\>,&
    e^{(79)} & = \left\{1,\tfrac{1}{z_7}\right\}\>,&
    e^{(579)} & = \left\{1,\tfrac{1}{z_5z_7}\right\}\>,&
\end{aligned}
\nonumber
\\
&\qquad
\begin{aligned}
    \qquad\qquad\qquad&e^{(2579)} = \left\{1,\tfrac{1}{z_2},\tfrac{1}{z_5 z_7},\tfrac{1}{z_2 z_5 z_7}\right\} , \ \\
    e  = &e^{(82579)} = \left\{1,\tfrac{1}{z_2 z_5 z_7},\tfrac{z_8}{z_2 z_5 z_7}\right\} ,
\end{aligned}
\end{align}
and as corresponding dual ones:
\begin{align}
&\qquad
\begin{aligned}
    h^{(9)} & = \left\{1\right\}\>,&
    h^{(79)} & = \left\{1,\delta_{7}\right\}\>,&
    h^{(579)} & = \left\{1,\delta_{57}\right\}\>,
\end{aligned}
\nonumber
\\
&\qquad
\begin{aligned}
    \qquad\qquad\qquad&h^{(2579)} = \left\{1,\delta_{2},\delta_{57},\delta_{257}\right\} , \ \\
    h = &h^{(82579)} = \left\{1,\delta_{257},z_8\delta_{257}\right\} .
\end{aligned}
\end{align}
We can then decompose the target left form: $\varphi \ =\ \frac{z_8^2}{z_2 z_5 z_7}$ in terms of the outer basis, where we omit the superscripts for simplicity, obtaining:
\begin{equation}
    \langle \varphi \vert \mathcal{C}] \ = \ c_4 \langle e_1 \vert \mathcal{C}] \ + \  c_{11} \langle e_2 \vert \mathcal{C}] \ + \  c_{12} \langle e_3 \vert \mathcal{C}]
    \>.
\end{equation}

\subsubsection*{Cut 1357, maximal cut of $J_5$}
Setting $z_1\ =\ z_3 \ = \ z_5 \ = \ z_7 \ = \ 0$, and choosing as order of variables, from outer to inner:
\begin{equation}
    \{ z_8 , \ z_2 , \ z_4 , \ z_6,\ z_9 \}
\end{equation}
we get as dimensions for the various layers:
\begin{equation}
    \nu\layer{9} \ = \ 1
    \>, \quad
    \nu\layer{69} \ = \ 2
    \>, \quad
    \nu\layer{469}\ = \ 2
    \>, \quad
    \nu\layer{2469} \ = \ 4
    \>, \quad
    \nu\layer{82469} \ = \ 4
    \>.
\end{equation}
We pick as bases:
\begin{align}
&\qquad\qquad
\begin{aligned}
    e^{(9)} & = \left\{1\right\}\>,&
    e^{(69)} & = \left\{1,\tfrac{1}{z_6}\right\}\>,&
    e^{(469)} & = \left\{1,\tfrac{1}{z_4z_6}\right\}\>,&
\end{aligned}
\nonumber
\\
&\qquad
\begin{aligned}
    \qquad\qquad\qquad&e^{(2469)} = \left\{1,\tfrac{1}{z_2},\tfrac{1}{z_4 z_6},\tfrac{1}{z_2 z_4 z_6}\right\} , \ \\
    e  = &e^{(82469)} = \left\{1,\tfrac{1}{z_2},\tfrac{1}{z_2 z_4 z_6},\tfrac{z_8}{z_2 z_5 z_6}\right\} ,
\end{aligned}
\end{align}
and as corresponding dual ones:
\begin{align}
&\qquad
\begin{aligned}
    h^{(9)} & = \left\{1\right\}\>,&
    h^{(69)} & = \left\{1,\delta_{6}\right\}\>,&
    h^{(469)} & = \left\{1,\delta_{46}\right\}\>,
\end{aligned}
\nonumber
\\
&\qquad
\begin{aligned}
    \qquad\qquad\qquad&h^{(2469)} = \left\{1,\delta_{2},\delta_{46},\delta_{246}\right\} , \ \\
    h = &h^{(82469)} = \left\{1,\delta_{2},\delta_{246},z_8\delta_{246}\right\} .
\end{aligned}
\end{align}
We can then decompose the target left form: $\varphi \ =\ \frac{z_8^2}{z_2 z_4 z_6}$ in terms of the outer basis, where we omit the superscripts for simplicity, obtaining:
\begin{equation}
    \langle \varphi \vert \mathcal{C}] \ = \ c_5 \langle e_1 \vert \mathcal{C}] \ + \  c_{9} \langle e_2 \vert \mathcal{C}] \ + \  c_{11} \langle e_3 \vert \mathcal{C}] \ + \  c_{12} \langle e_4 \vert \mathcal{C}]
    \>.
\end{equation}

\subsubsection*{Cut 2467, maximal cut of $J_6$}
Setting $z_2\ =\ z_4 \ = \ z_6 \ = \ z_7 \ = \ 0$, and choosing as order of variables, from outer to inner:
\begin{equation}
    \{ z_9 , \ z_1 , \ z_3 , \ z_5,\ z_8 \}
\end{equation}
we get as dimensions for the various layers:
\begin{equation}
    \nu\layer{8} \ = \ 1
    \>, \quad
    \nu\layer{58} \ = \ 2
    \>, \quad
    \nu\layer{358}\ = \ 4
    \>, \quad
    \nu\layer{1358} \ = \ 4
    \>, \quad
    \nu\layer{91358} \ = \ 4
    \>.
\end{equation}
We pick as bases bases:
\begin{align}
&\qquad\qquad
\begin{aligned}
    e^{(8)} & = \left\{1\right\}\>,&
    e^{(58)} & = \left\{1,\tfrac{1}{z_5}\right\}\>,&
    e^{(358)} & = \left\{1,\tfrac{1}{z_3},\tfrac{1}{z_5},\tfrac{1}{z_3 z_5}\right\}\>,&
\end{aligned}
\nonumber
\\
&\qquad\qquad\quad
\begin{aligned}
    \qquad\qquad\qquad&e^{(1358)} = \left\{1,\tfrac{1}{z_5},\tfrac{1}{z_1 z_3},\tfrac{1}{z_1 z_3 z_5}\right\} , \ \\
    e  = &e^{(91358)} = \left\{1,\tfrac{1}{z_5},\tfrac{1}{z_1 z_3 z_5},\tfrac{z_8}{z_1 z_3 z_5}\right\} ,
\end{aligned}
\end{align}
and as corresponding dual ones:
\begin{align}
&\qquad
\begin{aligned}
    h^{(8)} & = \left\{1\right\}\>,&
    h^{(58)} & = \left\{1,\delta_{5}\right\}\>,&
    h^{(358)} & = \left\{1,\delta_{3},\delta_{5},\delta_{35}\right\}\>,
\end{aligned}
\nonumber
\\
&\qquad\qquad
\begin{aligned}
    \qquad\qquad\qquad&h^{(1358)} = \left\{1,\delta_{5},\delta_{13},\delta_{135}\right\} , \ \\
    h = &h^{(91358)} = \left\{1,\delta_{5},\delta_{135},z_8\delta_{135}\right\} .
\end{aligned}
\end{align}
We can then decompose the target left form: $\varphi \ =\ \frac{z_8^2}{z_1 z_3 z_5}$ in terms of the outer basis, where we omit the superscripts for simplicity, obtaining:
\begin{equation}
    \langle \varphi \vert \mathcal{C}] \ = \ c_6 \langle e_1 \vert \mathcal{C}] \ + \  c_{10} \langle e_2 \vert \mathcal{C}] \ + \  c_{11} \langle e_3 \vert \mathcal{C}] \ + \  c_{12} \langle e_4 \vert \mathcal{C}]
    \>.
\end{equation}

\subsubsection*{Cut merging and symmetries}
From  the analysis on the complete spanning cuts, one is able to get the coefficients of the decomposition, obtaining:
\begin{align}
    &
    c_1 = c_2 = \frac{(3 d-10) (3 d-8) (s+2 t)}{(d-4)^2 (d-3) s^3}
    \>,\>
    c_3 = \frac{9 (3 d-10) (3 d-8)}{(d-4)^2 s t}
    \>,\>
    \nonumber \\
    &
    c_4 = \frac{2 (2 d s+2 d t-7 s-8 t)}{(d-4) s^2}
    \>,\>
    c_5 = \frac{9 (3 d-10)}{2 (d-4) s}
    \>,\>
    c_6 = \frac{(3 d-10) (2 s-t)}{(d-4) s^2}
    \>,\>
    \\
    &
    c_7 = c_8 = -\frac{(d-4) (7 s+9 t)}{2 (d-3) s}
    \>,\>
    c_9 = c_{10} = 4
    \>,\>
    c_{11} = \frac{(d-4) s t}{2 (d-3)}
    \>,\>
    c_{12} = -\frac{3 d s-12 s-2 t}{2 (d-3)}
    \nonumber
    \>.
\end{align}

Symmetries of the problem induce the following additional \textit{symmetry relations} between the master integrals:
\begin{equation}
    J_1 \ = \ J_2 \>, \qquad J_5 \ = \ J_6 \>, \qquad J_7 \ = \ J_8 \>, \qquad J_9 \ = \ J_{10}  \>.
\end{equation}
This reduce to 8 the number of genuinely independent master integrals, meaning that the final decomposition may be written as
\begin{align}
I = \tilde{c}_1 J_1 + c_3 J_3 + c_4 J_4 + \tilde{c}_5 J_5 + \tilde{c}_7 J_7 + \tilde{c}_9 J_9 + c_{11} J_{11} + c_{12} J_{12}
\>,
\end{align}
where
\begin{align}
\tilde{c}_1 = c_1+c_2\>, \qquad \tilde{c}_5 = c_5+c_6\>, \qquad \tilde{c}_7 =
    c_7+c_8\>, \qquad \tilde{c}_9 = c_9+c_{10}
    \>.
\end{align}

\subsection{Non-planar double-box}
\label{sec:nonplanardb}

\begin{figure}[H]
\centering
\includegraphics[width=0.85\textwidth]{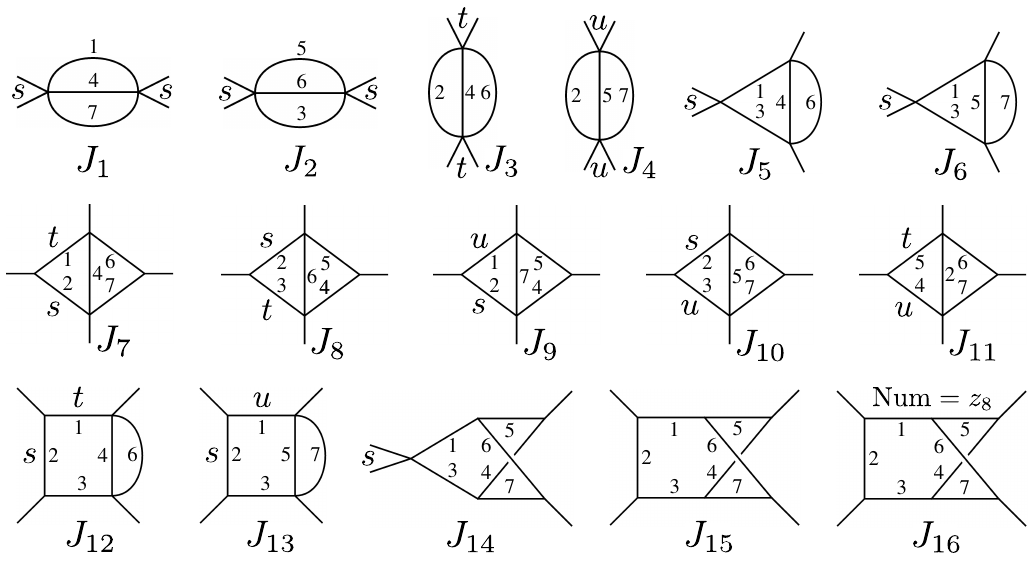}
\caption{The 16 master integrals for the non-planar double-box integral family, before imposing symmetry relations. The index $i$, next to the propagators, indicates the corresponding $z_i$ variable;
    $\mathrm{Num}$ stands for the numerator factor; $s, t,$ and $u$ channels are indicated, to distinguish among graphs with identical shape, but corresponding to different integrals.
\label{fig:npdoubleboxmasters}
}
\end{figure}

The integral family of the non-planar double box is given in terms of
\begin{align}
z_1 &= k_1^2\>, & z_2 &= (k_1{-}p_1)^2, & z_3 &= (k_1{-}p_1{-}p_2)^2, & z_4 &=
    (k_2{-}p_1{-}p_2{-}p_3)^2, & z_5 &= k_2^2\>,
    \nonumber \\
z_6 &= (k_1{-}k_2)^2, & z_7 &= (k_1{-}k_2{+}p_3)^2, & z_8 &= (k_1{-}p_1{-}p_2{-}p_3)^2, & z_9 &= (k_2{-}p_1)^2, &
\end{align}
where $z_8$ and $z_9$ are irreducible scalar products. The kinematics is such that:
\begin{equation}
    p_i^2 \ = \ 0 \>, \qquad s \ = \ (p_1{+}p_2)^2 \>, \qquad t \ = \ (p_1{+}p_4)^2 \>, \qquad s+t+u \ = \ 0 \>.
\end{equation}
This integral family has (before the application of symmetry relations) 16
master integrals, which we may pick as depicted in \figref{fig:npdoubleboxmasters}.
We are interested in decomposing the target integral:
\begin{equation}
    I  \ = \ \int \ \dd\mathbf{z} \  u(\mathbf{z}) \frac{z_8^2}{z_1z_2z_3z_4z_5z_6z_7}
    \label{eq:dbox_np_I}
\end{equation}
in terms of master integrals via a complete set of spanning cuts, as:
\begin{equation}
    I \ = \ \sum_{i=1}^{16} c_i \ J_i\>.
\end{equation}
The explicit expressions for the twist $u\brk{\mathbf{z}}$ as well as the master
integrals $J_i$ can be found in the ancillary
file~\filename{Dbox\_massless\_nonplanar.m}.

The set of spanning cuts is given by the maximal cuts
of the first six master integrals $\brc{J_1, \ldots, J_6}$, and we will now go through them
one by one.

\subsubsection*{Cut 147, maximal cut of $J_1$}
Setting $z_1\ =\ z_4 \ = \ z_7 \ = \ 0$, and choosing as order of variables, from outer to inner:
\begin{equation}
    \{ z_8 , \ z_3 , \ z_2 , \ z_5 , \ z_6,\ z_9 \}
\end{equation}
we get as dimensions for the various layers:
\begin{equation}
    \nu\layer{9} \ = \ 1
    \>, \quad
    \nu\layer{69} \ = \ 2
    \>, \quad
    \nu\layer{569} \ = \ 2
    \>, \quad
    \nu\layer{2569}\ = \ 4
    \>, \quad
    \nu\layer{32569} \ = \ 7
    \>, \quad
    \nu\layer{832569} \ = \ 6
    \>.
\end{equation}
We pick as bases:
\begin{eqnarray}
e^{(9)} & = & \left\{
1
\right\} ,
\qquad\qquad\quad
e^{(69)}  =  \left\{
1  ,   \tfrac{1}{z_6}
\right\} ,
\qquad\qquad\quad
e^{(569)}  =  \left\{
1 ,   \tfrac{1}{z_5z_6}
\right\} ,
\nonumber \\
e^{(2569)} & = & \left\{
1  ,   \tfrac{1}{z_2}  ,  \tfrac{1}{z_5 z_6}  ,  \tfrac{1}{z_2 z_5 z_6}
\right\} , \qquad
e^{(32569)}  =  \left\{
1 , \tfrac{1}{z_2}  ,  \tfrac{1}{z_6} ,  \tfrac{1}{z_2 z_6}  ,  \tfrac{1}{z_2 z_5 z_6} , \tfrac{1}{z_3 z_5 z_6} , \tfrac{1}{z_2 z_3 z_5 z_6}
\right\} , \nonumber \\
e&=&e^{(832569)} = \left\{
1  ,   \tfrac{1}{z_2 z_6}  , \tfrac{1}{z_2 z_5},
\tfrac{1}{z_3 z_5 z_6}, \tfrac{1}{z_2 z_3 z_5 z_6} ,  \tfrac{z_8}{z_2 z_3 z_5 z_6}
\right\} ,
\end{eqnarray}
and as corresponding dual ones:
\begin{eqnarray}
h^{(9)} \   & = & \  \left\{
1
\right\},  \qquad
h^{(69)} \   =  \  \left\{
1  ,   \delta_{6}
\right\} , \qquad
h^{(569)} \   =  \  \left\{
1  ,   \delta_{56}
\right\},  \
\nonumber\\
h^{(2569)} &  =  &  \left\{
1  ,   \delta_{2} ,  \delta_{56}  , \delta_{256}
\right\} , \qquad
h^{(32569)}   =    \left\{
1  ,  \delta_{2} , \delta_{6}, \delta_{26}  , \delta_{256}  , \delta_{356}, \delta_{2356}
\right\} , \
\nonumber\\
h&=&h^{(832569)} = \left\{
1  ,  \delta_{26} ,\delta_{25} , \delta_{356}  ,  \delta_{2356}, z_8 \delta_{2356}
\right\} \>.
\end{eqnarray}
We can then decompose the target left form: $\varphi \ =\ \frac{z_8^2}{z_2 z_3 z_5 z_6}$ in terms of the outer basis, where we omit the superscripts for simplicity, obtaining:
\begin{equation}
    \langle \varphi \vert \mathcal{C}] \ = \ c_1 \langle e_1 \vert \mathcal{C}] \ + \  c_7 \langle e_2 \vert \mathcal{C}] \ + \  c_{9} \langle e_3 \vert \mathcal{C}] \  + \  c_{14} \langle e_4 \vert \mathcal{C}] \ + \  c_{15} \langle e_5 \vert \mathcal{C}] \ + \  c_{16} \langle e_6 \vert \mathcal{C}]
    \>.
\end{equation}

\subsubsection*{Cut 356, maximal cut of $J_2$}
Setting $z_3\ =\ z_5 \ = \ z_6 \ = \ 0$, and choosing as order of variables, from outer to inner:
\begin{equation}
    \{ z_8 , \ z_1, \ z_2, \ z_7 , \ z_4,\ z_9 \}
\end{equation}
we get as dimensions for the various layers:
\begin{equation}
    \nu\layer{9} \ = \ 1
    \>, \quad
    \nu\layer{49} \ = \ 2
    \>, \quad
    \nu\layer{749} \ = \ 2
    \>, \quad
    \nu\layer{2749}\ = \ 4
    \>, \quad
    \nu\layer{12749} \ = \ 7
    \>, \quad
    \nu\layer{812749} \ = \ 6
    \>.
\end{equation}
We pick as bases:
\begin{eqnarray}
e^{(9)} & = & \left\{
1
\right\} ,
\qquad
e^{(49)}  =  \left\{
1  ,   \tfrac{1}{z_4}
\right\} ,
\qquad
e^{(749)}  =  \left\{
1 ,   \tfrac{1}{z_4z_7}
\right\} ,
\nonumber \\
e^{(2749)} & = & \left\{
1  ,   \tfrac{1}{z_2}  ,  \tfrac{1}{z_4 z_7}  ,  \tfrac{1}{z_2 z_4 z_7}
\right\} , \qquad
e^{(12749)}  =  \left\{
1  , \  \tfrac{1}{z_2}  ,  \tfrac{1}{z_4} ,  \tfrac{1}{z_2 z_4}  ,  \tfrac{1}{z_2 z_4 z_7} , \tfrac{1}{z_1 z_4 z_7} , \tfrac{1}{z_1 z_2 z_4 z_7}
\right\} , \nonumber \\
e&=&e^{(812749)} = \left\{
1  ,   \tfrac{1}{z_2 z_4}  , \tfrac{1}{z_2 z_7}, \tfrac{1}{z_1 z_4 z_7},  \tfrac{1}{z_1 z_2 z_4 z_7} ,  \tfrac{z_8}{z_1 z_2 z_4 z_7}
\right\} ,
\end{eqnarray}
and as corresponding dual ones:
\begin{eqnarray}
h^{(9)} \   & = & \  \left\{
1
\right\},  \qquad
h^{(49)} \   =  \  \left\{
1  ,   \delta_{4}
\right\} , \qquad
h^{(749)} \   =  \  \left\{
1  ,   \delta_{47}
\right\},  \
\nonumber\\
h^{(2749)} &  =  &  \left\{
1  ,   \delta_{2} ,  \delta_{47}  , \delta_{247}
\right\} , \qquad
h^{(12749)}   =    \left\{
1  ,  \delta_{2} , \delta_{4}, \delta_{24}  , \delta_{247}  , \delta_{147}, \delta_{1247}
\right\} , \
\nonumber\\
h&=&h^{(812749)} = \left\{
1  ,  \delta_{24} ,\delta_{27} , \delta_{147}  ,  \delta_{1247}, z_8 \delta_{1247}
\right\} \>.
\end{eqnarray}
We can then decompose the target left form: $\varphi \ =\ \frac{z_8^2}{z_1 z_2 z_4 z_7}$ in terms of the outer basis, where we omit the superscripts for simplicity, obtaining:
\begin{equation}
    \langle \varphi \vert \mathcal{C}] \ = \ c_2 \langle e_1 \vert \mathcal{C}] \ + \  c_8 \langle e_2 \vert \mathcal{C}] \ + \  c_{10} \langle e_3 \vert \mathcal{C}] \ + \  c_{14} \langle e_4 \vert \mathcal{C}] \ + \  c_{15} \langle e_5 \vert \mathcal{C}] \ + \  c_{16} \langle e_6 \vert \mathcal{C}]
    \>.
\end{equation}

\subsubsection*{Cut 246, maximal cut of $J_3$}
Setting $z_2\ =\ z_4 \ = \ z_6 \ = \ 0$, and choosing as order of variables, from outer to inner:
\begin{equation}
    \{ z_3 , \ z_8, \ z_1, \ z_7 , \ z_5,\ z_9 \}
\end{equation}
we get as dimensions for the various layers:
\begin{equation}
    \nu\layer{9} \ = \ 1
    \>, \quad
    \nu\layer{59} \ = \ 2
    \>, \quad
    \nu\layer{759} \ = \ 2
    \>, \quad
    \nu\layer{1759}\ = \ 6
    \>, \quad
    \nu\layer{81759} \ = \ 10
    \>, \quad
    \nu\layer{381759} \ = \ 7
    \>.
\end{equation}
We pick as bases:
\begin{eqnarray}
e^{(9)} & = & \left\{
1
\right\} ,
\qquad
e^{(59)}  =  \left\{
1  ,   \tfrac{1}{z_5}
\right\} ,
\qquad
e^{(759)}  =  \left\{
1 ,  \tfrac{1}{z_5z_7}
\right\} ,
\nonumber \\
e^{(1759)} & = & \left\{
1  ,   \tfrac{1}{z_1}  , \tfrac{1}{z_5},  \tfrac{1}{z_5 z_7}  , \tfrac{z_9}{z_5 z_7} , \tfrac{1}{z_1 z_5 z_7}
\right\} , \nonumber\\
e^{(81759)} & =  & \left\{
1 , \tfrac{1}{z_1} , \tfrac{1}{ z_5} , \tfrac{z_8}{z_5} , \tfrac{1}{z_7} , \tfrac{1}{z_1 z_7} , \tfrac{1}{z_5 z_7} , \tfrac{z_8}{z_5 z_7} , \tfrac{1}{z_1 z_5 z_7} , \tfrac{z_8}{z_1 z_5 z_7}
\right\} , \nonumber \\
e&=&e^{(381759)} = \left\{
1  ,   \tfrac{1}{z_1 z_7}  , \tfrac{1}{z_3 z_5},  \tfrac{1}{z_5 z_7},\tfrac{1}{z_1 z_3},  \tfrac{1}{z_1 z_3 z_5 z_7} ,  \tfrac{z_8}{z_1 z_3 z_5 z_7}
\right\}
\>,
\end{eqnarray}
and as corresponding dual ones:
\begin{eqnarray}
h^{(9)} \   & = & \  \left\{
1
\right\},  \qquad
h^{(59)} \   =  \  \left\{
1  ,   \delta_{5}
\right\} , \qquad
h^{(759)} \   =  \  \left\{
1  ,   \delta_{57}
\right\},  \
\nonumber\\
h^{(1759)} &  =  &  \left\{
1  ,   \delta_{1} ,  \delta_{5}  , \delta_{57} , z_9 \delta_{57}, \delta_{157}
\right\} , \nonumber \\
h^{(81759)}   & =   &  \left\{
1  ,  \delta_{1} , \delta_{5}, z_8 \delta_{5}  , \delta_{7}, \delta_{17}  , \delta_{57}, z_8 \delta_{57}, \delta_{157}, z_8 \delta_{157}
\right\} ,
\nonumber\\
h&=&h^{(381759)} = \left\{
1  ,  \delta_{17} ,\delta_{35} , \delta_{57}  , \delta_{13} , \delta_{1357}, z_8 \delta_{1357}
\right\}
\>.
\end{eqnarray}
We can then decompose the target left form: $\varphi \ =\ \frac{z_8^2}{z_1 z_3 z_5 z_7}$ in terms of the outer basis, where we omit the superscripts for simplicity, obtaining:
\begin{equation}
    \langle \varphi \vert \mathcal{C}] \ = \ c_3 \langle e_1 \vert \mathcal{C}] \ + \  c_7 \langle e_2 \vert \mathcal{C}] \ + \ c_{8} \langle e_3 \vert \mathcal{C}] \ + \ c_{11} \langle e_4 \vert \mathcal{C}] \ + \ c_{12} \langle e_5 \vert \mathcal{C}] \ + \ c_{15} \langle e_6 \vert \mathcal{C}] \ + \ c_{16} \langle e_7 \vert \mathcal{C}]
    \>.
\end{equation}

\subsubsection*{Cut 257, maximal cut of $J_4$}
Setting $z_2\ =\ z_5 \ = \ z_7 \ = \ 0$, and choosing as order of variables, from outer to inner:
\begin{equation}
    \{ z_1 , \ z_8, \ z_3, \ z_6 , \ z_4,\ z_9 \}
\end{equation}
we get as dimensions for the various layers:
\begin{equation}
    \nu\layer{9} \ = \ 1
    \>, \quad
    \nu\layer{49} \ = \ 2
    \>, \quad
    \nu\layer{649} \ = \ 2
    \>, \quad
    \nu\layer{3649}\ = \ 7
    \>, \quad
    \nu\layer{83649} \ = \ 10
    \>, \quad
    \nu\layer{183649} \ = \ 7
    \>.
\end{equation}
We pick as bases:
\begin{eqnarray}
e^{(9)} & = & \left\{
1
\right\} ,
\qquad
e^{(49)}  =  \left\{
1  ,   \tfrac{1}{z_4}
\right\} ,
\qquad
e^{(649)}  =  \left\{
1 ,  \tfrac{1}{z_4z_6}
\right\} ,
\nonumber \\
e^{(3649)} & = & \left\{
1  ,   z_3  , \tfrac{1}{z_6},  \tfrac{1}{z_3}  , \tfrac{1}{z_4 z_6} , \tfrac{z_3}{z_4z_6 },\tfrac{1}{z_3z_4z_6 }
\right\} , \nonumber\\
e^{(83649)} & =  & \left\{
1 , \tfrac{1}{z_3} , \tfrac{1}{ z_4} , \tfrac{z_3}{z_4} , \tfrac{1}{z_6} , \tfrac{1}{z_3 z_6}, \tfrac{1}{z_4 z_6} , \tfrac{z_8}{z_4 z_6}  , \tfrac{1}{z_3 z_4 z_6} , \tfrac{z_8}{z_3 z_4 z_6}
\right\} , \nonumber \\
e&=&e^{(183649)} = \left\{
1  ,   \tfrac{1}{z_1 z_4}  , \tfrac{1}{z_3 z_6},  \tfrac{1}{z_4 z_6},\tfrac{1}{z_1 z_3},  \tfrac{1}{z_1 z_3 z_4 z_6} ,  \tfrac{z_8}{z_1 z_3 z_4 z_6}
\right\}
\>,
\end{eqnarray}
and as corresponding dual ones:
\begin{eqnarray}
h^{(9)} \   & = & \  \left\{
1
\right\},  \qquad
h^{(49)} \   =  \  \left\{
1  ,   \delta_{4}
\right\} , \qquad
h^{(649)} \   =  \  \left\{
1  ,   \delta_{46}
\right\},  \
\nonumber\\
h^{(3649)} &  =  &  \left\{
1  ,  z_3 ,\delta_{6} ,  \delta_{3}  , \delta_{46} , z_3 \delta_{46}, \delta_{346}
\right\} , \nonumber \\
h^{(83649)}   & =   &  \left\{
1  ,  \delta_{3} , \delta_{4}, z_3 \delta_{4}  , \delta_{6}, \delta_{36}  , \delta_{46}, z_8 \delta_{46}, \delta_{346}, z_8 \delta_{346}
\right\} ,
\nonumber\\
h&=&h^{(183649)} = \left\{
1  ,  \delta_{14} ,\delta_{36} , \delta_{46}  , \delta_{13} , \delta_{1346}, z_8 \delta_{1346}
\right\}
\>.
\end{eqnarray}
We can then decompose the target left form: $\varphi \ =\ \frac{z_8^2}{z_1 z_3 z_4 z_6}$ in terms of the outer basis, where we omit the superscripts for simplicity, obtaining:
\begin{equation}
    \langle \varphi \vert \mathcal{C}] \ = \ c_4 \langle e_1 \vert \mathcal{C}] \ + \  c_9 \langle e_2 \vert \mathcal{C}] \ + \ c_{10} \langle e_3 \vert \mathcal{C}] \ + \ c_{11} \langle e_4 \vert \mathcal{C}] \ + \ c_{13} \langle e_5 \vert \mathcal{C}] \ + \ c_{15} \langle e_6 \vert \mathcal{C}] \ + \ c_{16} \langle e_7 \vert \mathcal{C}]
    \>.
\end{equation}

\subsubsection*{Cut 1346, maximal cut of $J_5$}
Setting $z_1\ =\ z_3 \ = \ z_4 \ = \ z_6 \ = \ 0$, and choosing as order of variables, from outer to inner:
\begin{equation}
    \{ z_8 , \ z_2 , \ z_5 , \ z_7,\ z_9 \}
\end{equation}
we get as dimensions for the various layers:
\begin{equation}
    \nu\layer{9} \ = \ 1
    \>, \quad
    \nu\layer{79} \ = \ 2
    \>, \quad
    \nu\layer{579}\ = \ 2
    \>, \quad
    \nu\layer{2579} \ = \ 4
    \>, \quad
    \nu\layer{82579} \ = \ 5
    \>.
\end{equation}
We pick as bases:
\begin{eqnarray}
e^{(9)} &=& \left\{
1
\right\} ,
\qquad
e^{(79)}  =  \left\{
1  ,   \tfrac{1}{z_7}
\right\} ,
\qquad
e^{(579)}  =  \left\{
1 ,   \tfrac{1}{z_5z_7}
\right\} ,
\nonumber \\
& &\quad\, e^{(2579)} = \left\{
1  ,   \tfrac{1}{z_2}  , \tfrac{1}{z_5 z_7} ,  \tfrac{1}{z_2 z_5 z_7}
\right\} ,
\nonumber \\
&e&=e^{(82579)} = \left\{
1  ,   \tfrac{1}{z_2}  , \tfrac{1}{z_5 z_7}, \tfrac{1}{z_2 z_5 z_7},  \tfrac{z_8}{z_2 z_5 z_7} \right\}
\>,
\end{eqnarray}
and as corresponding dual ones:
\begin{eqnarray}
h^{(9)} \   & = & \  \left\{
1
\right\},  \ \
h^{(79)} \   =  \  \left\{
1  ,   \delta_{7}
\right\} , \ \
h^{(579)} \   =  \  \left\{
1  ,   \delta_{57}
\right\},  \nonumber \\
& &h^{(2579)} =   \left\{
1  ,  \delta_{2} , \delta_{57}, \delta_{257}
\right\}\nonumber ,\\
h &=& h^{(82579)}  = \left\{
1  , \delta_{2}, \delta_{57}, \delta_{257}, z_8 \delta_{257}
\right\}
\>.
\end{eqnarray}
We can then decompose the target left form: $\varphi \ =\ \frac{z_8^2}{z_2 z_5 z_7}$ in terms of the outer basis, where we omit the superscripts for simplicity, obtaining:
\begin{equation}
    \langle \varphi \vert \mathcal{C}] \ = \ c_5 \langle e_1 \vert \mathcal{C}] \ + \  c_{12} \langle e_2 \vert \mathcal{C}] \ + \ c_{14} \langle e_3 \vert \mathcal{C}]  \ + \  c_{15} \langle e_4 \vert \mathcal{C}] \ + \ c_{16} \langle e_5 \vert \mathcal{C}]
    \>.
\end{equation}

\subsubsection*{Cut 1357, maximal cut of $J_6$}
Setting $z_1\ =\ z_3 \ = \ z_5 \ = \ z_7 \ = \ 0$, and choosing as order of variables, from outer to inner:
\begin{equation}
    \{ z_8 , \ z_2 , \ z_4 , \ z_6 , \ z_9 \}
\end{equation}
we get as dimensions for the various layers:
\begin{equation}
    \nu\layer{9} \ = \ 1
    \>, \quad
    \nu\layer{69} \ = \ 2
    \>, \quad
    \nu\layer{469}\ = \ 2
    \>, \quad
    \nu\layer{2469} \ = \ 4
    \>, \quad
    \nu\layer{82469} \ = \ 5
    \>.
\end{equation}
We pick as bases:
\begin{eqnarray}
e^{(9)} & = & \left\{
1
\right\} ,
\ \
e^{(69)}  =  \left\{
1  ,   \tfrac{1}{z_6}
\right\} ,
\ \
e^{(469)}  =  \left\{
1 ,   \tfrac{1}{z_4z_6}
\right\} ,
\nonumber \nonumber \\
& &\quad\, e^{(2469)}  =  \left\{
1  ,   \tfrac{1}{z_2}  , \tfrac{1}{z_4 z_6} ,  \tfrac{1}{z_2 z_4 z_6}
\right\} , \\
&e&=e^{(82469)}  =  \left\{
1  ,  \tfrac{1}{z_2},\tfrac{1}{z_4 z_6}, \tfrac{1}{z_2 z_4 z_6}  ,  \tfrac{z_8}{z_2 z_4 z_6} \right\} ,
\end{eqnarray}
and as corresponding dual ones:
\begin{eqnarray}
h^{(9)} \   & = & \  \left\{
1
\right\},  \ \
h^{(69)} \   =  \  \left\{
1  ,   \delta_{6}
\right\} , \ \
h^{(469)} \   =  \  \left\{
1  ,   \delta_{46}
\right\},  \nonumber \\
& &\quad h^{(2469)} = \left\{
1  ,   \delta_{2} , \delta_{46}, \delta_{246}
\right\} , \\
&h& = h^{(82469)}   =    \left\{
1  , \delta_{2},\delta_{46}, \delta_{246}, z_8 \delta_{246}
\right\} .
\end{eqnarray}
We can then decompose the target left form: $\varphi \ =\ \frac{z_8^2}{z_2 z_4 z_6}$ in terms of the outer basis, where we omit the superscripts for simplicity, obtaining:
\begin{equation}
    \langle \varphi \vert \mathcal{C}] \ = \ c_6 \langle e_1 \vert \mathcal{C}] \ + \  c_{13} \langle e_2 \vert \mathcal{C}] \ + \  c_{14} \langle e_3 \vert \mathcal{C}] \ + \  c_{15} \langle e_4 \vert \mathcal{C}]  \ + \  c_{16} \langle e_5 \vert \mathcal{C}]
    \>.
\end{equation}

\subsubsection*{Cut merging and symmetries}
From  the analysis on the complete spanning cuts, one is able to get the coefficients of the decomposition, obtaining:
\begin{align}
    c_1 & = c_2 = -\frac{(d-3) (3 d-10) (3 d-8) ((7 d-30) s+8 (2 d-9) t)}{4 (d-4)^3 (2 d-9) s^2 (s+t)}
    \>,
    \nonumber \\
    c_3 & = -\frac{3 (d-3) (3 d-10) (3 d-8) (4 (2 d-9) s+(5 d-22) t)}{2 (d-4)^3 (2 d-9) s t (s+t)}
    \>,
    \label{eq:csnpdbox} \\
    c_4 & = \frac{(d-3) (3 d-10) (3 d-8) (13 d s+9 d t-60 s-42 t)}{2 (d-4)^3 (2 d-9) s (s+t)^2}
    \>,\>\;
    c_5 = c_6 = -\frac{3 (d-3) (3 d-10)}{2 (d-4)^2 s}
    \>,
    \nonumber
    \\
    c_7 &= \frac{-11 d s-15 d t+48 s+66 t}{36 s-8 d s}
    \>,\>\;
    c_8 = \frac{-11 d s-15 d t+48 s+66 t}{36 s-8 d s}
    \>,\>\;
    c_9 = c_{10} = \frac{3 (14-3 d) t^2}{4 (2 d-9) s (s+t)}
    \>,\>
    \nonumber \\
    c_{11} & = -\frac{5 d s-4 d t-24 s+18 t}{2 (2 d-9) (s+t)}
    \>,\>\;
    c_{12} = c_{13} = -\frac{2 (d-3)}{d-4}
    \>,\>\;
    c_{14} = \frac{s}{4}
    \>,\>\;
    c_{15} = \frac{s t}{4}
    \>,\>\;
    c_{16} = \frac{1}{4} (2 t-3 s)
    \>.
    \nonumber
\end{align}
By applying symmetry relations one gets the following relations:
\begin{equation}
    J_1 \ = \ J_2 \>, \qquad J_5 \ = \ J_6 \>, \qquad J_7 \ = \ J_8 \>, \qquad J_{9} \ = \ J_{10} \>.
\end{equation}
that reduce to 12 the number of genuinely independent master integrals. A
reduction unto this basis is obtained by adding the corresponding coefficients
as given by eq.~\eqref{eq:csnpdbox}.

\subsection{Planar double-box with one external mass}
\label{subsec:1m_db_planar}
\begin{figure}[H]
\centering
\includegraphics[width=0.9\textwidth]{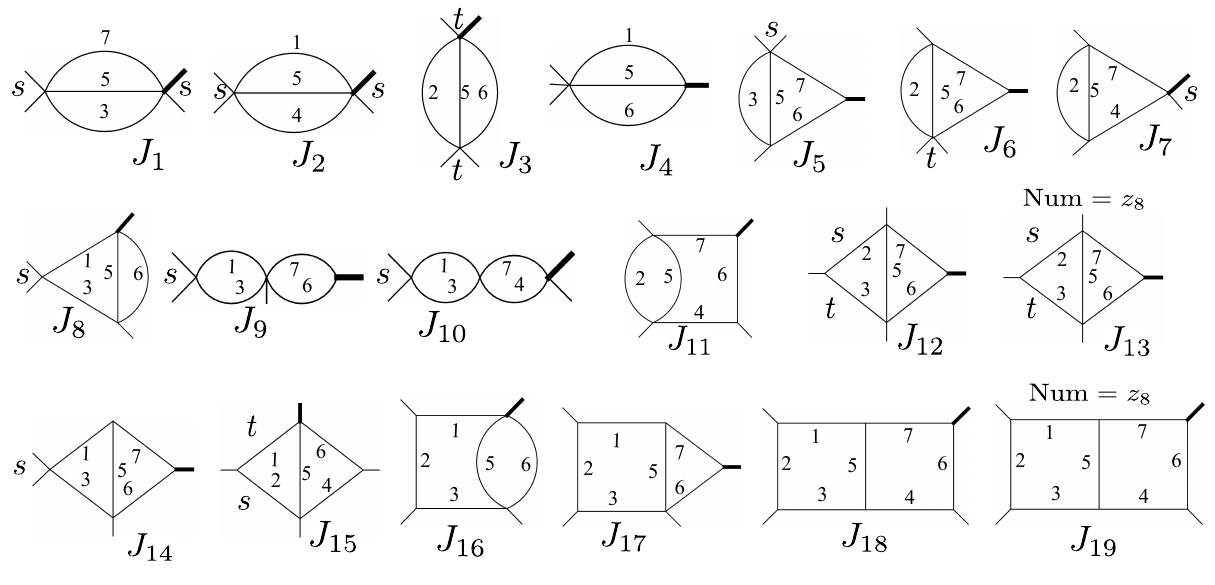}
\caption{The 19 master integrals of the planar double-box with one external mass, before imposing symmetry relations. The index $i$, next to the propagators, indicates the corresponding $z_i$ variable;
    $\mathrm{Num}$ stands for the numerator factor; $s, t,$ and $u$ channels are indicated, to distinguish among graphs with identical shape, but corresponding to different integrals.
    \label{fig:onemassdbplanar}
}
\end{figure}
The integral family of the planar double box with one external mass is given in terms of:
\begin{align}
    z_1 &= k_1^2\>, & z_2 &= (k_1{+}p_1)^2, & z_3 &= (k_1{+}p_1{+}p_2)^2, & z_4 &= (k_2{+}p_1{+}p_2)^2, & z_5 &= (k_1{-}k_2)^2,
    \nonumber \\
    z_6 &= (k_2{-}p_3)^2, & z_7 &= k_2^2\>, & z_8 &= (k_2{+}p_1)^2, & z_9 &= (k_1{-}p_3)^2, &
\end{align}
where $z_8$ and $z_9$ are ISPs. The kinematics is such that:
\begin{equation}
    p_3^2 = m^2 \>, \qquad p_1^2=p_2^2=p_4^2=0 \>,\qquad s = (p_1{+}p_2)^2 \>, \qquad t = (p_1{+}p_3)^2 \>, \qquad s{+}t{+}u = m^2 \>.
\end{equation}
This integral family has (before the application of symmetry relations) 19
master integrals, belonging to 17 sectors, which are depicted in~\figref{fig:onemassdbplanar}.
We are interested in decomposing the target integral:
\begin{equation}
    I  \ = \ \int \ \dd\mathbf{z} \  u(\mathbf{z}) \frac{z_8^2}{z_1z_2z_3z_4z_5z_6z_7}
\end{equation}
in terms of master integrals via a complete set of spanning cuts, as:
\begin{equation}
    I \ = \ \sum_{i=1}^{19} c_i \ J_i\>.
\end{equation}
The set of spanning cuts is given by the maximal cuts
of $\brc{J_1, \ldots, J_4,J_7,J_9,J_{10}}$.
The explicit expressions for the twist $u\brk{\mathbf{z}}$ as well as the master
integrals $J_i$ can be found in the ancillary
file \filename{Dbox\_1m\_planar.m}.

\subsection{Non-planar double-box with one external mass}
\label{subsec:1m_db_nonplanar_f2}
\begin{figure}[H]
\centering
\includegraphics[width=0.9\textwidth]{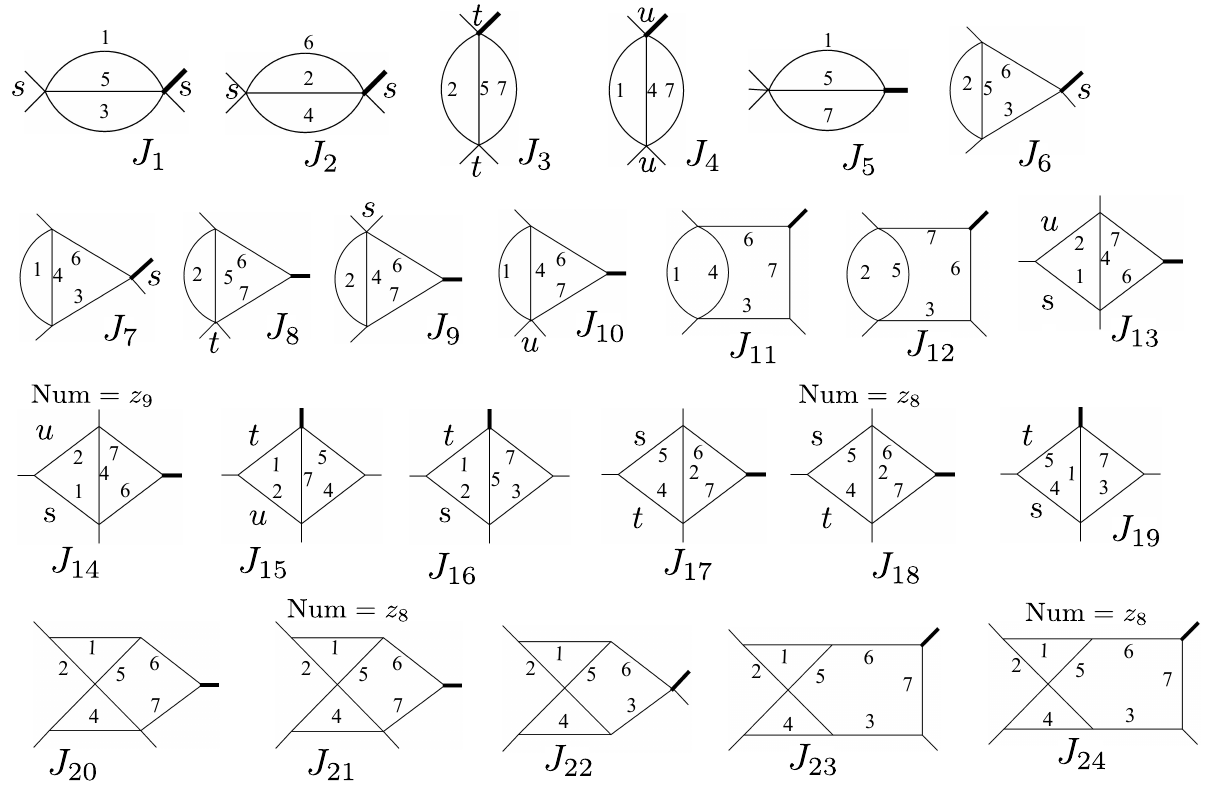}
\caption{
    The 24 master integrals of one of the non-planar double-box with one external mass, before imposing symmetry relations. The index $i$, next to the propagators, indicates the corresponding $z_i$ variable;
    $\mathrm{Num}$ stands for the numerator factor; $s, t,$ and $u$ channels are indicated, to distinguish among graphs with identical shape, but corresponding to different integrals.
}
    \label{fig:db_nonplanar_f2_sectors}
\end{figure}

This non-planar double-box with one external mass integral family is given in terms of:
\begin{align}
z_1 &= k_1^2\>, & z_2 &= (k_1{+}p_1)^2, & z_3 &= (k_2{+}p_1{+}p_2)^2, & z_4 &= (k_2{-}k_1{+}p_2)^2, & z_5 &= (k_2{-}k_1)^2, \nonumber \\
z_6 &= k_2^2\>, & z_7 &= (k_2{-}p_3)^2, & z_8 &= (k_2{+}p_1)^2, & z_9 &= (k_1{-}p_3)^2, &
\end{align}
where $z_8,z_9$ are ISPs. The kinematics is such that:
\begin{equation}
    p_3^2 = m^2 \>, \qquad p_1^2=p_2^2=p_4^2=0 \>,\qquad s = (p_1{+}p_2)^2 \>, \qquad t = (p_1{+}p_3)^2 \>, \qquad s{+}t{+}u = m^2 \>.
\end{equation}
This integral family has (before the application of symmetry relations) 24
master integrals, belonging to 20 sectors, which are depicted
in~\figref{fig:db_nonplanar_f2_sectors}. We are interested in decomposing the
target integral:
\begin{equation}
    I  \ = \ \int \ \dd\mathbf{z} \  u(\mathbf{z}) \frac{z_8^2}{z_1z_2z_3z_4z_5z_6z_7}
\end{equation}
in terms of master integrals via a complete set of spanning cuts, as:
\begin{equation}
    I \ = \ \sum_{i=1}^{24} c_i \ J_i\>.
\end{equation}

The set of spanning cuts is given by the maximal cuts
of $\brc{J_1, \ldots, J_7}$.
The explicit expressions for the twist $u\brk{\mathbf{z}}$ as well as the master
integrals $J_i$ can be found in the ancillary
file \filename{Dbox\_1m\_nonplanar.m}.\\

These applications represent significant milestones in the context of the complete decomposition of Feynman (two-loop) integrals in terms of master integrals by projection via intersection numbers, and, as such, they constitute another substantial part resulting from our work.

\section{Conclusions}
\label{sec:conclusions}

Intersection numbers are pivotal for investigation of the vector space formed
by twisted period integrals, influencing a broad spectrum of mathematical
and physical studies.
In this work, we have explored some new avenues of the twisted cohomology
theory that extend this framework and its range of applications,
which enabled us to propose an simplified version of the recursive algorithm
for evaluation of the intersection numbers for differential $n$-forms
\cite{Mizera:2019gea,Frellesvig:2019uqt}.

We have investigated the role of the evanescent regulators in the
computation of the intersection numbers within the framework of (ordinary) twisted
cohomology. These regulators, while being essential for the correct evaluation of
the intersection numbers in the traditional approach, may increase the complexity of the calculations.
Our careful analysis offered, on the one side, an independent, explicit proof that
the coefficients of the integral decomposition depend just on the leading term of the Laurent series
expansion in the regulator of the intersection numbers, and, on the other side, that
such a leading term can be computed directly within the relative twisted
cohomology theory, when using delta-forms as bases elements \cite{Caron-Huot:2021iev}.

Because of such established equivalence, we made use of the twisted relative cohomology to eliminate the need for
analytic regulators, and introduced a systematic algorithm for selecting
multivariate delta-forms as elements of the dual basis. This choice induces a block-triangular
structure in the metric and in the connection matrices,
therefore it simplifies the evaluation of the intersection numbers appearing in the master decomposition formula.

Additionally, we leveraged the polynomial division algorithm and the global
residue techniques \cite{Fontana:2023amt}, to bypass the need for algebraic
extensions and for polynomial factorization. In particular, we introduced a
novel polynomial ideal generator to simplify the recursive algorithm at each
stage of the sequence.

The simplified algorithm for the evaluation of intersection numbers between
$n$-forms, presented in this work, was successfully applied to the direct,
complete decomposition of two-loop, planar and non-planar, Feynman integrals,
that appear in the scattering amplitudes of either four massless particles or
three massless and one-massive particles.

Our theoretical investigation and the novel computing algorithm related to it,
constitute a significant progress in studying the algebraic properties of cohomology groups,
and their impact on the evaluation of Feynman integrals as well as on Euler-Mellin
integrals and Aomoto/Gelfand-Kapranov-Zelevinsky hypergeometric systems. These
contributions stand to enrich both the fields of physics and mathematics,
expanding our understanding of these intricate domains, and of their (still hidden)
connections.

We expect that the generalisation of the concepts discussed here,
within the context of the recursive approach to the evaluation of the intersection number,
such as polynomial divisions, global residues and delta-forms,
to the recently proposed algorithm based on Stokes' theorem in $n$ dimensions, and
that makes use of a single, higher-order partial differential equation
\cite{Chestnov:2022xsy}, therefore bypassing the need of fibrations, and, with
it, of the sequential iterations, can lead to the optimal computational
strategy for computing intersection numbers for $n$-forms. We defer such an
important evolution to be the subject of further studies.

\subsection*{Acknowledgements}
We wish to acknowledge Federico Gasparotto for insightful discussions on the theory of relative cohomology, important checks at various stages of the project, for comments on the manuscript,
and for collaboration at earlier stages.
We thank Gaia Fontana and Tiziano Peraro, for interesting discussions on 
their work \cite{Fontana:2023amt}, and for sharing with us their results before publication;
and Sid Smith, for interesting discussions on the evaluation of the intersection numbers, and for 
partial checks on the decomposition of the two-loop planar box diagram.
We acknowledge Sergio Cacciatori, Gaia Fontana, Yoshiaki Goto, Tiziano Peraro, Andrzej Pokraka, and Nobuki Takayama for comments on the manuscript.

V.C. is supported by the European Research Council (ERC) under the European
Union's research and innovation programme grant agreement 101040760 (ERC
Starting Grant FFHiggsTop), and in part by the Diagrammalgebra Stars Wild-Card
Grant UNIPD.
H.F. is supported by a Carlsberg Foundation Reintegration Fellowship, and has received funding from the European Union's Horizon 2020 research and innovation program under the Marie Sk{\l}odowska-Curie grant agreement No. 847523 'INTERACTIONS'.
The work of M.K.M is partially supported by Fellini~-~Fellowship for Innovation at INFN funded by the European Union's Horizon 2020 research and innovation programme under the Marie Sk{\l}odowska-Curie grant agreement No.~754496.

\newpage
\appendix
\section{Extended Euclidean Algorithm}
\label{app:eea}
The multiplicative inverse $\tilde{d}(z)$, of a polynomial $d(z)$, modulo
$\Baikov$, defined in eq.~\eqref{eq:mult_inverse}, can be obtained via the
Extended Euclidean Algorithm (EEA).
Given any two polynomials, say $a(z)$ and $b(z)$,
the EEA yields two polynomials, say $s(z) $ and $t(z)$
such that:
\begin{eqnarray}
    a(z) \, s(z) \ + \ b(z) \, t(z) \ = \  \GCD (a(z),b(z))
    \>,
\end{eqnarray}
where $\GCD(a(z), b(z))$ is the greatest common divisor of $a(z)$ and $b(z)$.

In our case, we can identify $a(z)=d(z)$, and $b(z)=\Baikov$, which, being
coprime, satisfy $\GCD(d(z),\Baikov)=1$. Therefore, the EEA gives:
\begin{eqnarray}
    d(z) \, s(z) \ + \ \Baikov \, t(z) \ = \  1 .
\end{eqnarray}
Reading the above equation modulo $\Baikov$ implies:
\begin{eqnarray}
    d(z) \, s(z) = \  1  \ \ mod \ \Baikov \ ,
\end{eqnarray}
hence, the function $s(z)$ is precisely the polynomial inverse%
\footnote{
    The {\sc Mathematica} function \code{PolynomialExtendedGCD} can be used to find the polynomial inverse.
}
of $d(z)$, i.e. $s(z) = \tilde{d}(z)$.

\section{Choice of Basis Elements: An Explicit Example}
\label{sec:pickingbases}
In this appendix we illustrate with an explicit example the procedure outlined in \secref{subsec:choice_of_basis_elements} for choosing basis elements.\\
~\\
Let us consider the three-mass elliptic sunrise integral. The propagators are given by
\begin{align}
    z_1 &= (k_1-p_1)^2 - m_1^2
    \>,
    &
    z_2 &= (k_1-k_2)^2 - m_2^2
    \>,
    &
    z_3 &= k_2^2 - m_3^2
    \>,
    \nonumber \\
    z_4 &= k_1^2
    \>,
    &
    z_5 &= (k_2-p_1)^2
    \>.
\end{align}
The variables $z_1, z_2, z_3$ are actual denominators, while $z_4$ and $z_5$ are irreducible numerator factors (ISPs). The kinematics is given by $p_1^2=s$.\\
~\\
The twist is given by
\begin{align}
    u = B^{\gamma}\>,
    \qquad \text{with} \qquad 
    \gamma = \frac{d-4}{2} \qquad \text{and} \qquad B = B_0 + B_1 + B_2
    \>,
\end{align}
where
\begin{align}
    B_0 &= \big( z_2{+}m_2^2{-}z_1{-}m_1^2{-}z_3{-}m_3^2{+}s \big) \left( (z_1{+}m_1^2)(z_3{+}m_3^2) - s (z_2{+}m_2^2) + z_4 z_5 \right)
    \>, \\[1mm]
    B_1 &= z_4 \left( (z_1{-}z_2{+}m_1^2{-}m_2^2) (z_3{+}m_3^2{-}s) + 2 (z_1{+}m_1^2) z_5 \right) - z_4^2 z_5
    \>, \\[1mm]
    B_2 &= B_1\big|_{
        z_1 \rightleftharpoons z_3
        \,,\,
        m_1 \rightleftharpoons m_3
        \,,\,
        z_4 \rightleftharpoons z_5
    }
    \>.
\end{align}
To generate a list of MIs we perform the following 4 steps:
\begin{enumerate}
    \item
        First, we list all the possible sectors
        \begin{equation}
            \bigbrc{\mathcal{S}_1, \dots, \mathcal{S}_8}
            =
            \bigbrc{
                \varnothing,
                \brc{z_1},
                \brc{z_2},
                \brc{z_3},
                \brc{z_1, z_2},
                \brc{z_1 , z_3},
                \brc{z_2 , z_3},
                \brc{z_1,z_2 , z_3}
            }
            \>,
            \label{eq:sectors}
        \end{equation}
        and initialize an empty list of MIs
        \begin{equation}
            \MIs=\varnothing
            \>.
        \end{equation}
        The counting of the number of zeros of $\omega_{\mathcal{S}_i}$ for $i=1, \dots, 4$ reads
        \begin{equation}
            \nu_{\mathcal{S}_i}=0, \qquad i =1, \dots, 4
            \>.
        \end{equation}
        This corresponds to the well-known fact that integrals of polynomials
        in loop momenta vanish in dimensional regularization, so no MIs are
        present.
    \item
        Next we consider $\mathcal{S}_5$ and observe
        \begin{equation}
           \nu_{\mathcal{S}_5}=1
            \>,
            \label{eq:omega_5_sunrise}
        \end{equation}
        hence we update the list of MIs with one element
        \begin{equation}
            \MIs= \left\{ \frac{1}{z_1 z_2} \right\}
            \>.
        \end{equation}
    \item
        Moving to $\mathcal{S}_6$ and $\mathcal{S}_7$, we find
        \begin{equation}
             \nu_{\mathcal{S}_6}=\nu_{\mathcal{S}_7}=1
             \>.
             \label{eq:omega_6_sunrise}
        \end{equation}
        The list of MIs at this stage receives two new elements and reads
        \begin{equation}
            \MIs= \left\{ \frac{1}{z_1 z_2},\frac{1}{z_1 z_3}, \frac{1}{z_2 z_3} \right\}.
        \end{equation}
    \item
        Finally, we analyze $\mathcal{S}_8$ having
        \begin{equation}
            \nu_{\mathcal{S}_8}=7
            \>.
            \label{eq:omega_8_sunrise}
        \end{equation}
        To avoid overcounting we need to subtract from
        eq.~\eqref{eq:omega_8_sunrise} contributions from all the
        subsectors. In particular, since $\mathcal{S}_i \subset \mathcal{S}_8$,
        for $i=5,6,7$, the corresponding $3$ MIs are already taken into account, so we
        only need to specify
        \begin{equation}
            \nu_{\mathcal{S}_8} - \sum_{i=5}^{7} \nu_{\mathcal{S}_i}=7-3=4
        \end{equation}
        new basis elements.
\end{enumerate}
Thus the full list of basis elements we constructed is
\begin{equation}
    \MIs=\left\{ \frac{1}{z_1 z_2}, \frac{1}{z_1 z_3}, \frac{1}{z_2 z_3}, \frac{1}{z_1 z_2 z_3}, \frac{z_4}{z_1 z_2 z_3}, \frac{z_5}{z_1 z_2 z_3}, \frac{z_5^2}{z_1 z_2 z_3} \right\}
\end{equation}
and the corresponding dual basis reads
\begin{align}
h = \left\{ \delta_{12}, \delta_{13}, \delta_{23}, \delta_{123}, z_4 \delta_{123}, z_5 \delta_{123}, z_5^2 \delta_{123} \right\}
\>.
\end{align}

The procedure to choose the bases at the intermediate layers of the iterative
algorithm is similar, the only difference is the choice of active
$z$-variables. For reader's convenience we collect the
dimensions of the cohomoloy groups
$\nu_{\cS_i}\supbbf{m}$ at each layer $\brk{\mathbf{m}}$ and in each sector
$\cS_i$ of eq.~\eqref{eq:sectors} in~\tabref{tab:sectors}.

\begin{table}
\begin{center}
    \setlength{\extrarowheight}{1mm}
    \NiceMatrixOptions{rules/color=[gray]{0.75}}
    \begin{NiceTabular}{ccccc|c}
        $\cS_i$ $\backslash$ $\brk{\mathbf{m}}$
    & $\brk{5}$
    & $\brk{35}$
    & $\brk{235}$
    & $\brk{1235}$
    & $\brk{41235}$
    \\
    $\varnothing$       & 1 & 1 & 0 & 0 & 0 \\ \Hline
    $\brc{z_1}$         & 1 & 1 & 0 & 0 & 0 \\ \Hline
    $\brc{z_2}$         & 1 & 1 & 1 & 1 & 0 \\ \Hline
    $\brc{z_3}$         & 1 & 2 & 1 & 1 & 0 \\ \Hline
    $\brc{z_1,z_2}$     & 1 & 1 & 1 & 2 & 1 \\ \Hline
    $\brc{z_1,z_3}$     & 1 & 2 & 1 & 2 & 1 \\ \Hline
    $\brc{z_2,z_3}$     & 1 & 2 & 3 & 3 & 1 \\ \Hline
    $\brc{z_1,z_2,z_3}$ & 1 & 2 & 3 & 6 & 7 \\
    \CodeAfter
        \tikz \draw [thick]  (1-|2) -- (10-|2);
        \tikz \draw [thick]  (2-|1) -- (2-|7);
\end{NiceTabular}
\end{center}
\caption{
    The cohomology dimensions $\nu\supbbf{m}_{\cS_i}$ defined in
    eq.~\protect\eqref{eq:leepom} for each sector~\protect\eqref{eq:sectors}
    labeling the rows, and on each fibration layer $\brk{\mathbf{m}}$ labeling
    the columns.
    The last column contains the counting numbers of
    eqs.~\protect\eqref{eq:omega_5_sunrise,eq:omega_6_sunrise,eq:omega_8_sunrise}.
    Columns in the middle collect the countings for the intermediate layers.
    \label{tab:sectors}
    }
\end{table}

\bibliographystyle{JHEP}
\bibliography{biblio}

\end{document}